\documentclass[%
 reprint,
nofootinbib,
 amsmath,amssymb,
 aps,
]{revtex4-2}

\usepackage[dvipsnames]{xcolor}
\usepackage{graphicx}
\usepackage{dcolumn}
\usepackage{bm}

\newcommand{\M}{m_{\text{P}}}

\begin{document}

\preprint{APS/123-QED}

\title{\boldmath Quintessential inflation in Palatini $f(R)$ gravity}

\author{Konstantinos Dimopoulos}
 \email{konst.dimopoulos@lancaster.ac.uk}
\author{Samuel S\'anchez L\'opez}%
 \email{s.sanchezlopez@lancaster.ac.uk}
\affiliation{%
 Consortium for Fundamental Physics, Physics Department,\\Lancaster University, Lancaster LA1 4YB, United Kingdom.
}%





\begin{abstract}
  We investigate in detail a family of quintessential inflation models in the
  context of $R+R^2$ Palatini modified gravity. We find that successful
  inflation and quintessence are obtained with an inflaton scalar potential
  that is approximately quadratic in inflation and inverse quartic in
  quintessence. We show that corrections for the kination period due to 
  Palatini modified gravity are subdominant, while the setup does not challenge
  constraints on modified gravity from solar system observations and microscopic
  experiments, in contrast to the metric case. We obtain concrete predictions
  regarding primordial tensors 
  to be probed in the near future.
\end{abstract}

\maketitle


\section{Introduction} 

Cosmic inflation is the most compelling solution to the fine-tuning problems
of hot big bang cosmology, while it simultaneously generates the
primordial curvature perturbation which is necessary for the formation of large
scale structure in the Universe \citep{Lyth:2009zz}.
Observations of ever increasing precision in
the last decades have confirmed the predictions of primordial inflation, while
the rival theory of structure formation (that of cosmic strings) has collapsed
\citep{Perivolaropoulos:2005wa}.

Meanwhile, about twenty years ago it was discovered that recently the Universe
has started engaging in another boot of late-time inflation
\citep{Riess:1998cb,Perlmutter:1998np}. 
This can be driven by a tiny positive value of the cosmological constant,
but the necessary fine-tuning is extreme (more than 120 orders of magnitude)
\citep{Weinberg:1988cp}. One prominent
alternative is using a scalar field to drive this late-time inflation, which has
been called quintessence \citep{Caldwell:1997ii};
the fifth element after baryons, cold dark matter,
photons and neutrinos. Employing a new dynamical degree of freedom to explain
late-time inflation introduces a new problem though, that of its initial
conditions. The problem was ameliorated by considering quintessence with
attractor properties that would lead to the desired late-time inflation with a
wide range of initial conditions \citep{Copeland:2006wr}.
However, recent precise observations seem to
undermine this kind of tracker quintessence. Therefore, another solution to the
problem of quintessence's initial conditions is necessary.

In their seminal paper \citep{Peebles:1998qn}, Peebles and Vilenkin considered
linking primordial to late-time inflation by using a single scalar field to
drive them both. Apart from being economic, this quintessential inflation
proposal fixed the initial conditions of quintessence through the inflationary
attractor. Quintessential inflation enables studying both early and late
inflation in a single theoretical framework. However, one needs to satisfy
simultaneously observations of the late and early Universe, hopefully without
introducing too many model parameters. The task is rather difficult as the
energy density in the two inflationary eras differs by more than a hundred
orders of magnitude. Moreover, new mechanisms for reheating the Universe had
to be devised, for the scalar field should not decay into the thermal bath of
the hot big bang, as with conventional inflation, but has to survive until the
present to play the role of quintessence.

Despite the challenging odds, a number
of successful quintessential inflation proposals have been constructed
(for a recent list of references see \citep{Dimopoulos:2017zvq,Dimopoulos:2017tud,Hossain:2014zma,Geng:2015fla,Benisty:2020qta,Akrami:2017cir,Akrami:2020zxw}). In many
of them the scalar potential features two flat regions: the inflationary plateau
and the quintessential tail. The flat regions are connected through a steep part
of the potential, which, when traversed by the scalar field, results in a boost
of kinetic energy density, which dominates the Universe. This period, called
kination \citep{Joyce:1997fc},
is a unique prediction of quintessential inflation and leads to
characteristic observational signatures such as a spike in the spectrum of
primordial gravitational waves \citep{Sahni:2001qp},
which may be observed by the forthcoming LISA
mission. During kination, the inflaton-quintessence field is oblivious of the
scalar potential (because it is dominated by its kinetic energy density), which
means that the predictions of kination are independent of the particular
quintessential inflation scalar potential considered.


Apart from employing a scalar field with a suitable potential in Einstein
gravity, inflation can be also achieved by suitably modifying gravity, as was
discovered early on by Starobinsky, in his seminal paper
\citep{Starobinsky:1980te}, where he
introduced a higher-order term in the gravitational action, schematically
$R+R^2$, where $R$ is the scalar curvature (Ricci scalar). Even though it is
possible to model primordial inflation is this way, a~la Starobinsky inflation
or Higgs inflation \citep{Bezrukov:2007ep}
for example, the task is much harder for late-time
inflation. Indeed, many attempts to consider modified gravity theories, e.g.
with a term proportional to $1/R$ in the gravitational action \citep{Carroll:2003wy},
were shown to be unstable\footnote{There are still many other viable models of $f(R)$ gravity in the metric formalism that successfully generate late-time acceleration, such as $f(R)=R-\mu R^p$ with $p\in (0,1)$, originally proposed in \citep{Amendola:2006we}.} \citep{Sawicki:2007tf}.
Moreover, the recent observation confirming that the speed of
propagation of gravitational waves is exactly light-speed (to precision of 15
orders of magnitude) \citep{Monitor:2017mdv}, as
suggested by Einstein gravity, excludes the contemplation of many otherwise
motivated modifications of gravity at work in the late Universe (for example
the Gauss-Bonnet term \citep{vandeBruck:2017voa}). While work still continues
in this front \citep{Nojiri:2017ncd},
in this paper we have investigated a blended quintessential inflation model,
which achieves primordial inflation via modified gravity (also called $f(R)$
gravity), but late-time inflation via a suitable scalar potential of
quintessence. However, our approach cannot be clean-cut in that modifications
of gravity are expected to affect the kination era, after primordial inflation,
and the recent history of the Universe, after the end of the radiation era
(when $R\simeq 0$). 

One complication we had to face was that in the Starobinsky $R+R^2$ model, the
higher order gravity term introduces an extra degree of freedom, which can be
rendered in the form of a scalar field, the scalaron. Starobinsky inflation is
very successful, but if it were to be considered as part of a quintessential
inflation model, the scalaron would need to survive until today and become
quintessence. In this case though, experimental tests of gravity \citep{Hoyle:2004cw} cannot allow
successful primordial inflation.
Therefore, we considered a different version of
gravity, where the $R+R^2$ model does not introduce a scalaron and the theory
does not conflict with the experimental tests of gravity. In this version,
called Palatini gravity (originally introduced by Einstein \citep{Ferraris_1982}), both the metric and the connection are
independent dynamical degrees of freedom (gravitational fields).
The inflaton field is not the scalaron (for the latter does not exist) but it is
explicitly introduced, as in conventional inflation.

However, Palatini gravity does affect our scenario.
Firstly, it ``flattens'' the inflaton scalar potential
\citep{Antoniadis:2018ywb,Antoniadis:2018yfq,Lloyd-Stubbs:2020pvx,Gialamas:2020snr,Gialamas:2019nly}
so that the desired inflationary plateau can be attained even with an originally
steep scalar potential. Secondly, the theory is expected to introduce
modifications to the kination period, after primordial inflation, and also
in the late Universe, when the inflaton field becomes quintessence. We
investigate in detail what these effects are, considering a family of models,
which is a generalised version of the original Peebles-Vilenkin quintessential
inflation potential.

Our paper is structured as follows. In Sec.~II, we provide a brief pedagogic
description of Palatini $f(R)$ gravity. In Sec.~III. we introduce $R+R^2$
Palatini gravity with a scalar field and background matter/radiation, with
emphasis on the interaction between them. In Sec~IV, we present our family of
quintessential inflation models and how they are affected by the assumed
modified gravity setup, focusing on the period of primordial inflation.
In Sec.~V, the period of kination in the context of Palatini $R+R^2$ gravity
is investigated, in a way which is independent on the form of the scalar
potential. To obtain concrete inflationary predictions we assume gravitational
reheating, but our results are easy to reproduce when considering another, more
efficient mechanism, as only the relevant number of inflationary e-folds is
affected. In Section~VI, we investigate quintessence in our setup and look for
the amount of tuning needed to satisfy the coincidence requirement.
In Sec.~VII, we show how experimental gravity tests are not challenged by the
modified gravity theory considered. Finally, we end in Sec.~VIII with a brief
discussion of our findings and our conclusions.

We use natural units with \mbox{$c=\hbar=1$} and 
$\M=(8\pi G)^{-1/2}=2.43\times 10^{18} \text{GeV}$ being 
the reduced Plank mass. The signature of the metric is
diag($-$1,+1,+1,+1).

\section{\boldmath Palatini $f(R)$ gravity}\label{aoruogabiuwef}

Before we fix the potential $V(\varphi)$ for our quintessential inflation model and solve the inflationary, kination and quintessential dynamics, we introduce some general concepts in relation to $f(R)$ theories of modified gravity.

$f(R)$ theories are defined by the action
\begin{eqnarray}
    S=\frac{1}{2}\M^2\int\text{d}^4x\sqrt{-g}f(R)+S_m[g_{\mu\nu},\Phi],
    \label{aofgfaiubwer}
\end{eqnarray}
where $S_m$ is the matter action and $\Phi$ collectively represents the matter fields. In the present work the matter action is taken to be 
\begin{eqnarray}
    S_m[g_{\mu\nu},\Phi]&&=\int\text{d}^4x\sqrt{-g}\left[-\frac{1}{2}g^{\mu\nu}\nabla_{\mu}\varphi\nabla_{\nu}\varphi-V(\varphi)\right]\nonumber\\
    &&+S_m[g_{\mu\nu},\psi],
\end{eqnarray}
where $\varphi$ is the inflaton, $V(\varphi)$ its potential and $\psi$ collectively represents all the matter fields other than the inflaton. 

The action \eqref{aofgfaiubwer} is dynamically equivalent to 
\begin{eqnarray}
  S&&=\frac12 {\M^2}
  \int\text{d}^4x\sqrt{-g}\left[f(\chi)+f'(\chi)(R-\chi)\right]\nonumber\\
    &&+S_m[g_{\mu\nu},\Phi],
    \label{afoigfjainerg}
\end{eqnarray}
where we have introduced the auxiliary field $\chi$. Indeed, the equation of motion for $\chi$ (if $f''(\chi)\neq 0$) leads to $\chi=R$, which can be replaced in Eq. \eqref{afoigfjainerg} to obtain the original action.

There exist two different approaches in order to obtain the field equations. Firstly, in the $\textit{metric}$ formalism, as in general relativity (GR), 
the metric $g_{\mu\nu}$ is taken to be the only independent gravitational field, \textit{i.e.}, the connection takes the usual Levi-Civita form of $g_{\mu\nu}$, which $\textit{a priori}$ is  not necessary. The equations of motion are given by
\begin{eqnarray}
    \frac{\delta S}{\delta g^{\mu\nu}}=0.
\end{eqnarray}

Secondly, in the $\textit{Palatini}$ formalism, both the connection $\Gamma^{\alpha}_{\mu\nu}$ and the metric $g_{\mu\nu}$ are taken to be independent gravitational fields, \textit{i.e.}, the connection does not necessarily take the Levi-Civita form of $g_{\mu\nu}$, although the matter action does not depend on the independent connection. The Riemann tensor $R^{\alpha}_{\phantom{\beta}\beta\mu\nu}=\partial_{\mu}\Gamma^{\alpha}_{\nu\beta}+\Gamma^{\alpha}_{\mu\lambda}\Gamma^{\lambda}_{\nu\beta}-(\mu\leftrightarrow\nu)$ and the Ricci tensor $R_{\mu\nu}\equiv R^{\rho}_{\phantom{\mu}\mu\rho\nu}$ are functions of the connection only, while the Ricci scalar $R\equiv g^{\mu\nu}R_{\mu\nu}$ also depends on the metric. The equations of motion are given by
\begin{eqnarray}
    \frac{\delta S}{\delta g^{\mu\nu}}=0\quad \text{and} \quad \frac{\delta S}{\delta \Gamma ^{\lambda}_{\mu\nu}}=0.
    \label{aorgaiuwbefibawef}
\end{eqnarray}

One obviously expects that the dynamics obtained from the action \eqref{aofgfaiubwer} departs from GR in both formalisms (although they both reduce to GR when $f(R)=R$). However, the way in which they do is different. In the metric formalism, a new dynamical degree of freedom, named the scalaron, given by
\begin{eqnarray}
s \equiv \frac{\text{d} f(R)}{\text{d} R}\equiv f_R,
\end{eqnarray}
is introduced. Its dynamical evolution is determined by the trace of the equation of motion in the Jordan frame
\begin{eqnarray}
    f_R(R)R-2f(R)+3\square f_R(R)=\frac{1}{\M^2}T.
\end{eqnarray}

It is also easy to see that metric $f(R)$ gravity is equivalent to a Brans-Dicke theory with parameter $\omega=0$ \citep{Sotiriou:2008rp}. This means that in the Einstein frame, after an appropriate field redefinition, the new dynamical degree of freedom appears in the action with a canonical kinetic term and minimally coupled to gravity. In sharp contrast to this scenario we have the Palatini formalism, where no new dynamical degree of freedom is introduced. The trace of the metric equation of motion now reads
\begin{eqnarray}
    f_R(R)R-2f(R)=\frac{1}{\M^2}T,
    \label{aworugbiqwe}
\end{eqnarray}
and it follows that $R$, and therefore $f(R)$ and $f_R(R)$, is algebraically related to the matter sources through the trace of the energy momentum tensor. Using the Brans-Dicke representation of the theory it is easy to see that Palatini $f(R)$ gravity is equivalent to a Brans-Dicke theory with parameter $\omega=-3/2$ \citep{Sotiriou:2008rp}. Thus, in the Einstein frame $s=f_R$ has no kinetic term. The way in which the gravitational dynamics is modified in the Palatini formalism is through the introduction of new effective matter sources (see below).

After clarifying one of the most important distinctions between both approaches to $f(R)$ modified theories of gravity we focus solely on the Palatini formalism. The equations of motion \eqref{aorgaiuwbefibawef} read\footnote{Eq. \eqref{aoweurbhubawer} has been obtained after some elementary manipulations and under the assumption that the connection is symmetric in its lower indices ($\textit{a priori}$ there is no need for this).}\citep{Sotiriou:2008rp}
\begin{eqnarray}
    f_{R}R_{(\mu\nu)}-\frac12 {f}g_{\mu\nu}=\frac{1}{\M^2}T_{\mu\nu},
    \label{aodbgibawe}
\end{eqnarray}
\begin{eqnarray}
    \nabla_{\lambda}(\sqrt{-g}f_{R}g^{\mu\nu})=0,
    \label{aoweurbhubawer}
\end{eqnarray}
where $R_{(\mu\nu)}$ represents the symmetric part of the Ricci tensor and the energy-momentum tensor is obtained in the usual way
\begin{eqnarray}
    T_{\mu\nu}=-\frac{2}{\sqrt{-g}}\frac{\delta S}{\delta g^{\mu\nu}}.
\end{eqnarray}

A couple of comments are in order. Firstly, just as in GR, the action \eqref{aofgfaiubwer} is invariant under diffeomorphisms. Thus, the conservation of the energy-momentum tensor, \textit{i.e.}, $\nabla_{\mu}T^{\mu\nu}=0$, is always satisfied. In Ref. \citep{Koivisto:2005yk}, it was shown that the energy-momentum tensor is also conserved in more general theories of modified gravity, with non-minimal couplings between gravity and the matter Lagrangian. It follows that the conservation equation for a perfect fluid in a FRW universe
\begin{eqnarray}
    \Dot{\rho}+3H(\rho+p)=0
\end{eqnarray}
is unchanged with respect to GR.
 
Secondly, under a conformal transformation $g_{\mu\nu}\rightarrow \bar{g}_{\mu\nu}=f_Rg_{\mu\nu}$, Eq. \eqref{aoweurbhubawer} reads
 \begin{eqnarray}
     \nabla_{\lambda}(\sqrt{-\bar{g}}\bar{g}^{\mu\nu})=0.
 \end{eqnarray}
 
 This equation can be solved algebraically to obtain the Levi-Civita connection of $\bar{g}_{\mu\nu}$
 \begin{eqnarray}
     \Gamma^{\alpha}_{\mu\nu}=\frac{1}{2}\bar{g}^{\alpha\beta}(\partial_{\mu}\bar{g}_{\beta\nu}+\partial_{\nu}\bar{g}_{\beta\mu}-\partial_{\beta}\bar{g}_{\mu\nu}).
     \label{aigbiawer}
 \end{eqnarray}

Given that, from Eq. \eqref{aworugbiqwe}, $R$ is algebraically related to $T$, and since we have an expression for $\Gamma^{\alpha}_{\mu\nu}$ in terms of $R$ and $g^{\mu\nu}$ (just plug $\bar{g}_{\mu\nu}=f_{R}g_{\mu\nu}$ in Eq. \eqref{aigbiawer}) we can eliminate the independent connection from the equations of motion and express them only in terms of the matter fields and the metric. Indeed, after some algebra, we obtain\citep{Sotiriou:2008rp}
\begin{eqnarray}
    G_{\mu\nu}&&=R_{\mu\nu}(g)-\frac{1}{2}g_{\mu\nu}R(g)\nonumber
    \\
    &&=\frac{1}{\M^2f_{R}}T_{\mu\nu}-\frac{1}{2}g_{\mu\nu}\left(R-\frac{f}{f_{R}}\right)\nonumber
    \\
    &&+\frac{1}{f_{R}}(\nabla_{\mu}\nabla_{\nu}-g_{\mu\nu}\square)f_{R}\nonumber
    \\
    &&-\frac{3}{2f_{R}^2}\left[\nabla_{\mu}f_{R}\nabla_{\nu}f_{R}-\frac{1}{2}g_{\mu\nu}(\nabla f_{R})^2\right],
    \label{aofjgbaiwner}
\end{eqnarray}
where $R_{\mu\nu}(g)$, $R(g)$ and $\nabla_{\mu}\nabla_{\nu}f_{R}$ are computed in terms of the Levi-Civita connection of the metric $g_{\mu\nu}$, while $R$ and $f_{R}$ must be seen as functions of $T$. 

It is important to note that in vacuum $T_{\mu\nu}=0$ the solution of Eq. \eqref{aworugbiqwe} is a constant\footnote{Unless \mbox{$f\propto R^2$}.} $R=R_{\rm vac}$, so that $f_{R}=f_{R_{\rm vac}}$ is also a constant. In this way Eq. \eqref{aofjgbaiwner} is reduced to 
\begin{eqnarray}
    G_{\mu\nu}=-\Lambda_{\text{eff}}g_{\mu\nu},
    \label{argbaiwebf}
\end{eqnarray}
where 
\begin{eqnarray}
    \Lambda_{\text{eff}}\equiv\frac{1}{2}\left(R-\frac{f}{f_{R}}\right)\Bigr|_{R_{\text{vac}}}=\frac14 {R_{\text{vac}}}
    \label{aowrgiabuwef}
\end{eqnarray}
plays the role of an effective cosmological constant and we have used Eq. \eqref{aworugbiqwe} with $T=0$ in the second step.

It is also possible to express the Einstein equations in terms of the metric $\bar{g}_{\mu\nu}=f_Rg_{\mu\nu}$ (below it is explained that this metric is the one corresponding to the Einstein frame). They read \citep{Olmo:2011uz}
\begin{eqnarray}
    G_{\mu\nu}(h)=\frac{1}{\M^2f_R}T_{\mu\nu}-\Lambda(T)\bar{g}_{\mu\nu},
    \label{aisdubfasdf}
\end{eqnarray}
where
\begin{eqnarray}
    \Lambda(T)=\frac{R f_R-f}{2f_R^2},
    \label{aodbfasdf}
\end{eqnarray}
where $R$ and $f_R$ are functions of the matter content, as explained above.

\section{The Model}

We work in $f(R)$ gravity with a Starobinski term, as in \citep{Antoniadis:2018ywb,Antoniadis:2018yfq,Lloyd-Stubbs:2020pvx}. In this way, we have
\begin{eqnarray}
    f(R)=R+\frac{\alpha}{2\M^2}R^2,
    \label{aorauiefibnwaef}
\end{eqnarray}
so that the action reads
\begin{eqnarray}
    S&&=\int\text{d}^4x\sqrt{-g}\bigg[\frac12{\M^2}R+\frac14{\alpha}R^2\nonumber\\
    &&-\frac{1}{2}g^{\mu\nu}\nabla_{\mu}\varphi\nabla_{\nu}\varphi-V(\varphi)\bigg]+S_m[g_{\mu\nu},\psi].
    \label{aijbgaweref}
\end{eqnarray}

As a remark, although $S_m[g_{\mu\nu},\psi]=0$ during inflation, we keep this term explicit in what follows since the treatment is also valid for the kination and quintessential sector of the theory, when matter and radiation fields are present. 

It is straightforward to calculate the matter dependence of the Ricci scalar. The derivatives of the $f(R)$ function read
\begin{eqnarray}
    f_R(R)=1+\frac{\alpha}{\M^2}R
    \label{audfgbdfaa}
\end{eqnarray}
and
\begin{eqnarray}
    f_{RR}(R)=\frac{\alpha}{\M^2}.
    \label{aourugbiuwbef}
\end{eqnarray}

Taking into account the two main contributions to the  energy density of the Universe come from the inflaton (or quintessence, depending on the cosmological era under consideration) and from regular pressureless mater and radiation, the energy momentum tensor can be written as, assuming the background matter and radiation behave as a perfect fluid,
\begin{eqnarray}
    T_{\mu\nu}^{\text{tot}}=T_{\mu\nu}^{(\varphi)}+T^{\rm B}_{\mu\nu},
\end{eqnarray}
where 
\begin{eqnarray}
    T_{\mu\nu}^{(\varphi)}=\partial_{\mu}\varphi\partial_{\nu}\varphi-g_{\mu\nu}\left[\frac{1}{2}\partial^{\alpha}\varphi\partial_{\alpha}\varphi+V(\varphi)\right]
\end{eqnarray}
and 
\begin{eqnarray}
    T^{\rm B}_{\mu\nu}=(\rho+p)u_{\mu}u_{\nu}+p g_{\mu\nu},
    \label{adofabsdf}
\end{eqnarray}
where $u^{\mu}$ is the four-velocity of a comoving observer with respect to the fluid (so that $-1=\eta^{\mu\nu}u_{\mu}u_{\nu}$).

It follows that the trace of the energy momentum tensor, remembering $\varphi=\varphi(t)$, reads
\begin{eqnarray}
    T^{\text{tot}}=T^{(\varphi)}+T^{\rm B},
\end{eqnarray}
where 
\begin{eqnarray}
    T^{(\varphi)}&&=g^{\mu\nu}T^{(\varphi)}_{\mu\nu}=-\partial^{\alpha}\varphi\partial_{\alpha}\varphi-4V(\varphi)\nonumber\\
    &&=\Dot{\varphi}^2-4V(\varphi),
\end{eqnarray}
where in the last equation we have taken $\varphi$ as homogeneous, and 
\begin{eqnarray}
    T^{\rm B}&&=g^{\mu\nu}T^{\rm B}_{\mu\nu}
    =-\rho+3p=-\rho(1-3w),
    \label{asd9ufbaiusdf}
\end{eqnarray}
where $\rho$ and $p$ are the density and the pressure of the background perfect fluid respectively, and \mbox{$w\equiv p/\rho$} is its barotropic parameter.

Assuming the Universe is filled up with radiation ($w_{\text{r}}=1/3$) and pressureless matter ($w_{\text{m}}=0$) we have
\begin{eqnarray}
    T^{\rm B}=-\rho_{\text{m}}.
\end{eqnarray}

Thus, Eq. \eqref{aworugbiqwe} reads 
\begin{eqnarray}
    f_RR-2f&&=\left(1+\frac{\alpha}{\M^2}R\right)R-2R-\frac{\alpha}{\M^2}R^2=-R\nonumber\\
    &&=\frac{1}{\M^2}\left(-\rho_{\text{m}}+\Dot{\varphi}^2-4V(\varphi)\right).
\end{eqnarray}

The curvature scalar is then obtained as a function of the matter content of the Universe as 
\begin{eqnarray}
    R=\frac{1}{\M^2}\left(\rho_{\text{m}}-\Dot{\varphi}^2+4V(\varphi)\right).
    \label{oabrgribaweqq}
\end{eqnarray}

Depending on the cosmological era under consideration, some approximations can be made to simplify Eq.~\eqref{oabrgribaweqq}. During slow-roll inflation $\rho_{\text{m}}=0$ and $\Dot{\varphi}^2\ll V(\varphi)$ so that the Ricci scalar reads
\begin{eqnarray}
    R_{\text{SR}}=\frac{4}{\M^2}V(\varphi).
    \label{aoguuaubwdf}
\end{eqnarray}

During kination, remembering the inflaton is kinetically dominated $\Dot{\varphi}^2\gg V(\varphi)$ and the other contribution to the energy-momentum tensor is radiation, which is traceless, we have
\begin{eqnarray}
    R_{\text{kin}}=-\frac{\Dot{\varphi}^2}{\M^2}.
    \label{aorugbaibwef}
\end{eqnarray}

During the radiation dominated era, $\rho_{\text{r}}\gg \rho^{(\varphi)}=\frac{1}{2}\dot{\varphi}^2+V(\varphi)$. However, since the energy-momentum tensor of a perfect fluid with $w=1/3$ is traceless, we have
\begin{equation}
    R_{\text{RD}}=\frac{1}{\M^2}(-\dot{\varphi}^2+4V(\varphi)).
\end{equation}
At reheating, the moment at which radiation becomes the dominant component in the Universe, the field is still in free fall as during kination (see below), so that we still have $R_{\text{RD}}\simeq -\dot{\varphi}^2/\M^2$. However, not long after reheating, the field stops and freezes (see Eq. \eqref{aoregbiawef} below), so that $R_{\text{RD}}\simeq 4V(\varphi)/\M^2$, which is extremely small since \mbox{$V\sim 10^{-120}m_{\rm P}^4$} because of the coincidence requirement (see below).

During the matter dominated era, the Ricci scalar reads
\begin{equation}
    R_{\text{MD}}=\frac{\rho_{\text{m}}}{\M^2}.
\end{equation}
Note also that during this era the energy density of the quintessence field is $\rho^{(\varphi)}\simeq V(\varphi)$ since it stops its roll-down the potential and freezes during the radiation dominated era, as explained above. 

During the quintessence era, the Ricci scalar still obeys Eq. \eqref{oabrgribaweqq}. However, we consider thawing quintessence (see below), which means the inflaton is only starting to unfreeze today, so that 
\begin{eqnarray}
    R_{\text{quin}}=\frac{1}{\M^2}\left(\rho_{\text{m}}+4V(\varphi)\right).
    \label{aodbfibdsf}
\end{eqnarray}

Finally, in vacuum, where $T_{\mu\nu}=0$, we have
\begin{eqnarray}
    R_{\text{vac}}=0.
    \label{afouabwibdfwef}
\end{eqnarray}

It is interesting that Palatini $f(R)$ gravity with a Starobinski term does not lead to gravity-driven inflation \citep{Meng:2003bk}, as in its metric $f(R)$ counterpart. As we have commented above, Palatini $f(R)$ theories do not introduce a new degree of freedom and the change in the gravitational dynamics (compared to conventional GR) can be interpreted as a change in the matter sources. In this way, when $\rho_{\text{m}}=\rho_{\text{r}}=0$, we re-obtain the conventional Friedmann equation with $H=0$, and inflation does not take place in the absense of an inflaton field. If we do introduce a minimally coupled scalar field $\varphi$ in Palatini $f(R)$ gravity with a Starobinski term, the standard inflationary dynamics (in the Jordan frame) is not affected when the inflaton is in the slow-roll regime \citep{Meng:2004yf} at the level of the background evolution. However, the generation of perturbations, which is behind the inflationary observables, is indeed affected.

Following the procedure through which we obtained the action in Eq.~\eqref{afoigfjainerg}, the action in Eq.~\eqref{aijbgaweref} is dynamically equivalent to
\begin{eqnarray}
    S&&=\int\text{d}^4x\sqrt{-g}\bigg[\frac12{\M^2}\left(1+\frac{\alpha}{\M^2}\chi\right)R-\frac14{\alpha}\chi^2\nonumber\\
    &&-\frac{1}{2}(\nabla\varphi)^2-V(\varphi)\bigg]+S_m[g_{\mu\nu},\psi].
    \label{ijabrgibar}
\end{eqnarray}

We emphasize that the original action in Eq. \eqref{aijbgaweref} can be obtained by imposing the constraint on the auxiliary field 
\begin{eqnarray}
    \frac{\delta S}{\delta \chi}=0
    \label{aodjdgfnijarng}
\end{eqnarray}
in the action in Eq. \eqref{ijabrgibar}.

We now perform a conformal transformation\footnote{As opposed to $f(R)$ gravity in the metric formalism, the Ricci tensor now only depends on the connection, so that it does not transform under the conformal transformation in Eq. \eqref{aiffgjbaierg}.}
\begin{eqnarray}
    g_{\mu\nu}\rightarrow\bar{g}_{\mu\nu}=f'(\chi)g_{\mu\nu}=\left(1+\frac{\alpha}{\M^2}\chi\right)g_{\mu\nu},
    \label{aiffgjbaierg}
\end{eqnarray}
so that 
\begin{eqnarray}
        &&\text{d}\bar{t}=\sqrt{f'(\chi)}\text{d}t\nonumber\\&&\bar{a}(\bar{t})=\sqrt{f'(\chi)}a(t).
        \label{aiudsbfbiasdf}
\end{eqnarray}

After some algebra, the action in the Einstein frame can be found to be
\begin{eqnarray}
    S&&=\int \text{d}^4x\sqrt{-\bar{g}}\Bigg[\frac12{\M^2}\bar{R}-\frac{1}{2}\frac{\M^2(\bar{\nabla}\varphi)^2}{(\M^2+\alpha\chi)}\nonumber\\
    &&-\frac{\M^4(V(\varphi)+\frac{\alpha}{4}\chi^2)}{(\M^2+\alpha\chi)^2}\Bigg]+S_m[(f'(\chi))^{-1}\bar{g}_{\mu\nu},\psi],
    \label{aoijdfbgiargr}
\end{eqnarray}
where barred quantities are calculated using the Einstein frame metric given by Eq. \eqref{aiffgjbaierg}. Note the new coupling between $\chi$ and the matter fields in the matter action. Now, imposing the condition in Eq. \eqref{aodjdgfnijarng} on the auxiliary field $\chi$, we have
\begin{eqnarray}
    \chi=\frac{4V(\varphi)+(\bar{\nabla}\varphi)^2}{\M^2-\frac{\alpha}{\M^2}(\bar{\nabla}\varphi)^2},
\end{eqnarray}
which implies that
\begin{eqnarray}
    f'(\chi(\varphi))=\frac{\M^4+4\alpha V(\varphi)}{\M^4-\alpha(\bar{\nabla}\varphi)^2}.
    \label{aoibgiawef}
\end{eqnarray}

Substituting back in the action in Eq. \eqref{aoijdfbgiargr}, one obtains 
\begin{eqnarray}
    &&S=\int\text{d}^4x\sqrt{-\bar{g}}\Bigg[\frac12{\M^2}\bar{R}-\frac{\frac{1}{2}(\bar{\nabla}\varphi)^2}{1+\frac{4\alpha}{\M^4}V(\varphi)}-\frac{V(\varphi)}{1+\frac{4\alpha}{\M^4}V(\varphi)}\nonumber
    \\
    &&+\frac{\alpha}{4\M^4}\frac{(\bar{\nabla}\varphi)^2(\bar{\nabla}\varphi)^2}{1+\frac{4\alpha}{\M^4}V(\varphi)}\Bigg]+S_m\left[(f'(\varphi))^{-1}\bar{g}_{\mu\nu},\psi\right],
    \label{aoiodfgjbairg}
\end{eqnarray}
where $(f'(\varphi))^{-1}$ is given by Eq.~\eqref{aoibgiawef} and the prime denotes a derivative with respect to $\chi=\chi(\varphi)$.

Since we study the behaviour of the inflaton during slow-roll and of quintessence today, when its potential is becoming shallow, higher than quadratic powers of $\bar{\nabla}\varphi$ are not expected to play a role. Furthermore, it can be shown \citep{Meng:2004yf} that during a kinetic energy dominated era, such a kination, the kinetic energy of the inflaton (in the Jordan frame) is bounded as 
\begin{eqnarray}
    \frac{1}{2}\Dot{\varphi}^2<\frac{\M^4}{2\alpha}.
    \label{aodufbaidsf}
\end{eqnarray}

As it is shown below, during kination the kinetic term in the action is canonical to a very good approximation (since the potential is negligible compared to $\M^4$ during this epoch). This means the canonical field in the Einstein frame $\phi$ is equal to the canonical field in the Jordan frame $\varphi$. Therefore, Eq. \eqref{aodufbaidsf} holds in the Einstein frame during kination and the quartic kinetic term in Eq. \eqref{aoiodfgjbairg} is negligible compared to the quadratic kinetic term, in the same way as it is during slow-roll inflation and during the quintessence tail. Thus, this term is ignored in what follows.

\subsection{\label{aidbfbiahsdf}Coupling to Matter}

The conformal transformation in Eq.~\eqref{aiffgjbaierg} introduces a coupling between the field $\varphi$ and the matter action in the Einstein frame, as can be seen in, \textit{e.g.}, Eq.~\eqref{aoiodfgjbairg}. In this section we investigate the effects of such coupling. 

The relation between the energy-momentum tensor in the Jordan and in the Einstein frames reads (remember barred quantities correspond to the Einstein frame while unbarred quantities correspond to the Jordan frame)
\begin{eqnarray}
    \bar{T}^{\rm B}_{\mu\nu}&&=-\frac{2}{\sqrt{-\bar{g}}}\frac{\delta S_m}{\delta \bar{g}^{\mu\nu}}=-\frac{2}{\sqrt{-\bar{g}}}\frac{\partial g^{\alpha\beta}}{\partial \bar{g}^{\mu\nu}}\frac{\delta S_m}{\delta g^{\alpha\beta}}\nonumber\\
    &&=\frac{f'(\varphi)}{(f'(\varphi))^2}\left(-\frac{2}{\sqrt{-g}}\frac{\delta S_m}{\delta g^{\mu\nu}}\right)=\frac{1}{f'(\varphi)}T^{\rm B}_{\mu\nu},
    \label{adsiufubauisdfa}
\end{eqnarray}
where we have used 
\begin{eqnarray}
    \frac{\partial g^{\alpha\beta}}{\partial \bar{g}^{\mu\nu}}=f'(\varphi)\delta^{\alpha}_{\mu}\delta^{\beta}_{\nu}
    \label{adofabisdf}
\end{eqnarray}
and
\begin{eqnarray}
    \sqrt{-\bar{g}}=(f'(\varphi))^2\sqrt{-g},
    \label{aosdjbfasdf}
\end{eqnarray}
which follow from Eq. \eqref{aiffgjbaierg}. Following \citep{Carroll:2003wy,Magnano:1993bd}, it is then convenient to define the energy-momentum tensor for a perfect fluid in the Einstein frame as 
\begin{eqnarray}
    \bar{T}^{\rm B}_{\mu\nu}=(\bar{\rho}+\bar{p})\bar{u}_{\mu}\bar{u}_{\nu}+\bar{p}\bar{g}_{\mu\nu},
\end{eqnarray}
where, comparing with Eq. \eqref{adofabsdf} and using Eq. \eqref{adsiufubauisdfa}, 
\begin{eqnarray}
        &&\bar{u}_{\mu}=\sqrt{f'(\varphi)}u_{\mu},\nonumber\\
        &&\bar{\rho}=\frac{\rho}{(f'(\varphi))^2},\nonumber\\
        &&\bar{p}=\frac{p}{(f'(\varphi))^2}.
        \label{aodoufabusdf}
\end{eqnarray}

During inflation, $S_m[g_{\mu\nu},\psi]=0$, and so the new coupling in the matter action between $\varphi$ and the matter fields does not change the dynamics. However, after inflation ends and the Universe is reheated, the matter action is not zero anymore. Indeed, the equation of motion for the inflaton field now reads
\begin{eqnarray}
    \frac{\delta S}{\delta \varphi}+\frac{\delta S_m}{\delta \varphi}=0,
\end{eqnarray}
where the result of the first term depends on the specific form the potential takes. 

Let us investigate the second term. We have
\begin{eqnarray}
  &&\frac{\delta S_m}{\delta \varphi}=\frac{\partial g^{\mu\nu}}{\partial \varphi}\frac{\delta S_m}{\delta g^{\mu\nu}}=f'_\varphi(\varphi)
  \bar{g}^{\mu\nu}\left(-\frac12{\sqrt{-g}}T^{\rm B}_{\mu\nu}\right)\nonumber\\
    &&=\frac{f'_\varphi(\varphi)}{f'(\varphi)}\bar{g}^{\mu\nu}\left(-\frac12{\sqrt{-\bar{g}}}\bar{T}^{\rm B}_{\mu\nu}\right)=-\frac{f'_\varphi(\varphi)}{2f'(\varphi)}\sqrt{-\bar{g}}\bar{T}^{\rm B},
    \label{adofuuahusdf}
\end{eqnarray}
where \mbox{$f'_\varphi=\partial f'/\partial\varphi$} (recall that the prime denotes derivative with respect to $\chi(\varphi)$) and we have used Eqs.~\eqref{aiffgjbaierg} and \eqref{adsiufubauisdfa}-\eqref{aosdjbfasdf}.

Analogously to Eq.~\eqref{asd9ufbaiusdf}, the trace of the energy-momentum tensor in the Einstein frame reads 
\begin{eqnarray}
  \bar{T}^{\rm B}&&=\bar{g}^{\mu\nu}\bar{T}^{\rm B}_{\mu\nu}=
  -\bar{\rho}+3\bar{p}=-\bar{\rho}(1-3\bar{w}),
\end{eqnarray}
where \mbox{$\bar{w}=\bar p/\bar\rho$} is the barotropic parameter of the background perfect fluid. Note that Eq. \eqref{aodoufabusdf} implies the barotropic parameter is the same in both the Jordan and Einstein frames. Indeed,
\begin{eqnarray}
    \bar{w}=\frac{\bar{p}}{\bar{\rho}}=\frac{p}{\rho}=w.
\end{eqnarray}
Furthermore, the prefactor in the right-hand-side of Eq.~\eqref{adofuuahusdf} reads, from Eq. \eqref{aoibgiawef} 
\begin{eqnarray}
  \frac{f'_\varphi(\varphi)}{f'(\varphi)}
    &&=\frac{4\alpha}{\M^4}\frac{\partial V}{\partial\varphi}\frac{1}{1+\frac{4\alpha}{\M^4} V(\varphi)}.
\end{eqnarray}

Putting everything together, we finally have
\begin{eqnarray}
    \frac{\delta S_m}{\delta \varphi}=\sqrt{-\bar{g}}\frac{2\alpha}{\M^4}\frac{\partial V(\varphi)}{\partial \varphi}\frac{\bar{\rho}(1-3\bar{w})}{1+\frac{4\alpha}{\M^4} V(\varphi)}.
    \label{adofbbasdf}
\end{eqnarray}

It immediately follows that during the radiation dominated epoch ($\bar{w}=1/3$)
\begin{eqnarray}
    \frac{\delta S_m}{\delta \varphi}\Bigr|_{\text{RD}}=0,
    \label{aaosdijfbasdf}
\end{eqnarray}
and the dynamics of $\varphi$ is unaffected by the new coupling in the matter action. Likewise, during kination, although the dominant contribution to the energy density of the Universe is that of the inflaton and the barotropic parameter of the Universe is $\bar{w}=1$, the only other matter field present during this epoch is radiation, so that $\bar{w}=1/3$ in Eq. \eqref{adofbbasdf} and the dynamics of the inflaton during kination is also unaffected.

As a remark, below is defined a new canonical field $\phi$ which is identified as the inflaton. Obtaining its equation of motion
\begin{eqnarray}
    \frac{\delta S}{\delta \phi}+\frac{\delta S_m}{\delta \phi}=0
    \label{aosdbfbiuasdf}
\end{eqnarray}
is straightforward by simply using the chain rule
\begin{eqnarray}
    \frac{\delta S_m}{\delta \phi}=\frac{\text{d} \varphi}{\text{d}\phi}\frac{\delta S_m}{\delta \varphi}.
    \label{adofbasdfasdf}
\end{eqnarray}

Finally, it has been explained above (see Eqs. \eqref{aisdubfasdf}-\eqref{aodbfasdf}) that the Einstein equations in the Einstein frame read 
\begin{eqnarray}
    \bar{G}_{\mu\nu}=\frac{1}{\M^2 f'(T)}T_{\mu\nu}-\frac{R(T) f'(T)-f(T)}{2(f'(T))^2}\bar{g}_{\mu\nu},
    \label{aoijdbfaiusdf}
\end{eqnarray}
where the Ricci scalar and the function $f'(R)$ depend on the matter content of the specific cosmological epoch under consideration.

\section{The Inflationary Sector}

After the general treatment of the action given in the previous sections, we fix $V(\varphi)$ to be the generalised quintessential inflation potential (the original model was proposed by P. J. E. Peebles and A. Vilenkin \citep{Peebles:1998qn}) given~by
\begin{eqnarray}
    V(\varphi)&&=\frac{\lambda^n}{\M^{n-4}}(\varphi^n+M^n), \quad \varphi<0\nonumber\\
    &&=\frac{\lambda^n}{\M^{n-4}}\frac{M^{n+q}}{\varphi^q+M^q},\quad \varphi\geq 0,
    \label{aouguan}
\end{eqnarray}
where $\lambda$ is a dimensionless constant fixed by the inflationary observables and \mbox{$0<M\ll \M$} is a suitable energy scale that is fixed by requiring that the potential energy density of the inflaton (see below) at its frozen value $\phi_F$ corresponds to the vacuum energy density measured today (coincidence requirement). The parameters $n$ and $q$ are of order unity. We will consider integer values of $n$ and $q$ to facilitate our analytic treatment, but this is stricly speaking not necessary, as we elaborate in the discussion section. The original potential of Ref.~\cite{Peebles:1998qn} is recovered when \mbox{$n=q=4$}.
Remember, as we have said above, that $S_m[g_{\mu\nu},\psi]=0$ during inflation.

The kinetic term in the action \eqref{aoiodfgjbairg}, when $|\varphi|\gg M$, reads
\begin{eqnarray}
    \frac{\frac{1}{2}(\bar{\nabla}\varphi)^2}{1+\frac{4\alpha}{\M^4}V(\varphi)}\simeq\frac{\frac{1}{2}(\bar{\nabla}\varphi)^2}{1+\frac{4\alpha\lambda^n}{\M^n}\varphi^n}.
    \label{aodbabiwdf}
\end{eqnarray}

It can be made canonical by means of the transformation
\begin{eqnarray}
    \text{d}\phi=\frac{\text{d}\varphi}{\sqrt{1+\frac{4\alpha\lambda^n}{\M^n}\varphi^n}}=\frac{\M}{\lambda(4\alpha)^{1/n}}\frac{\text{d}x}{\sqrt{1+x^n}},
    \label{aourubgiuawef}
\end{eqnarray}
where we have defined 
\begin{eqnarray}
    x\equiv \frac{\lambda(4\alpha)^{1/n}\varphi}{\M }
\end{eqnarray}
and $\phi$ can be identified as the canonical inflaton. For now it is not necessary to obtain $\phi=\phi(x)$. We only need 
\begin{eqnarray}
    \frac{\text{d}x}{\text{d}\phi}=\frac{\lambda(4\alpha)^{1/n}}{\M}\sqrt{1+x^n}.
    \label{oaubgbiuawef}
\end{eqnarray}

The potential in the Einstein frame reads 
\begin{equation}
    \bar{V}=\frac{V(\varphi)}{1+\frac{4\alpha}{\M^4}V(\varphi)}=\frac{\lambda^n\varphi^n/\M^{n-4}}{1+\frac{4\alpha\lambda^n}{\M^n}\varphi^n}=\frac{\M^4}{4\alpha}\frac{x^n(\phi)}{1+x^n(\phi)}.
    \label{aruguauiwebn}
\end{equation}

The slow-roll parameters are calculated in terms of the canonical field $\phi$, so that
\begin{eqnarray}
  \epsilon_V&&=\frac12{\M^2}\left(\frac{\bar{V}'(\phi)}{\bar{V}(\phi)}\right)^2=\frac12\frac{\M^2}{\bar V^2(x)}\left(\frac{\text{d}x}{\text{d}\phi}
  \frac{\partial \bar{V}(x)}{\partial x}\right)^2\nonumber\\
    &&=\frac12{\lambda^2(4\alpha)^{2/n}n^2}\frac{1}{x^2(1+x^n)},
    \label{oiabwergib}
\end{eqnarray}
where from now on the prime denotes derivative with respect to $\phi$, and 
\begin{eqnarray}
    \eta_V&&=\M^2\frac{\bar{V}''(\phi)}{\bar{V}(\phi)}=\frac{\M^2}{\bar{V}(x)}\frac{\text{d}x}{\text{d}\phi}\frac{\text{d}}{\text{d}x}\left(\frac{\text{d}x}{\text{d}\phi}\frac{\partial \bar{V}(x)}{\partial x}\right)\nonumber\\
    &&=\lambda^2(4\alpha)^{2/n}\frac{n(n-1)-n(\frac{n}{2}+1)x^n}{x^2(1+x^n)},
    \label{aijrbghubaweeee}
\end{eqnarray}
where we have used Eq. \eqref{oaubgbiuawef}. We can now calculate the remaining number of inflationary e-folds after the cosmological scales exit the horizon as
\begin{eqnarray}
    N&&=-\frac{1}{\M}\int_{\phi_{*}}^{\phi_{\text{end}}}\frac{\text{d}\phi}{\sqrt{2\epsilon_V(\phi)}}=-\frac{1}{\lambda^2n(4\alpha)^{2/n}}\int_{x_{*}}^{x_{\text{end}}}\text{d}x\nonumber\\
    &&=\frac{1}{2\lambda^2n(4\alpha)^{2/n}}(x_{*}^2-x_{\text{end}}^2)
    \label{a9eubuawe},
\end{eqnarray}
where $\phi_{*}$ is the inflaton value at which the cosmological scales leave the horizon and $\phi_{\text{end}}$ is the inflaton value at which inflation ends, \textit{i.e.}, $\epsilon_V(\phi_{\text{end}})=1$. The value of the field $x$ at the end of inflation $x_{\text{end}}\equiv x(\phi_{\text{end}})$ can be obtained, using Eq. \eqref{oiabwergib}, through the condition
\begin{eqnarray}
    \epsilon_V(\phi_{\text{end}})=1\Leftrightarrow x_{\text{end}}^2(x_{\text{end}}^n+1)=\frac12{\lambda^2(4\alpha)^{2/n}n^2}.
    \label{aoidfaijwbef}
\end{eqnarray}

For the typical values of $\lambda$ and $\alpha$ we consider, and for $n$ not too large, we have
\begin{eqnarray}
    |x_{\text{end}}|\ll 1
    \label{adofasiudfhasdf}
\end{eqnarray}
so that\footnote{In the opposite limit $|x_{\text{end}}|\gg1$, the term $n/4$ in the parenthesis in Eq. \eqref{oajbjregbiw} is replaced by the complicated expression $2^{\frac{-n-4}{n+2}}n^{\frac{-n+2}{2+n}}(\lambda^2(4\alpha)^{2/n})^{\frac{-n}{2+n}}$. Using the limit $|x_{\text{end}}|\gg 1$ it can be shown this expression is bounded from above by $n/4$. Taking into account that $N\gg n$ for reasonable values of $n$, this means that our results are insensitive to whether $|x_{\text{end}}|\gg1$ or $|x_{\text{end}}|\ll1$. However, we emphasize that, for the typical values of $\lambda$ and $\alpha$, $|x_{\text{end}}|\ll1$ holds.}
\begin{eqnarray}
    x_{\text{end}}^2\simeq \frac12{\lambda^2(4\alpha)^{2/n}n^2}\quad \text{for all $n$}.
    \label{aogiubwefd}
\end{eqnarray}

The value of the field $x$ when the cosmological scales leave the horizon
\mbox{$x_{*}=x(\phi_*)$} then reads, from Eq.~\eqref{a9eubuawe},
\begin{equation}
    x_{*}^2=x_{\text{end}}^2+2\lambda^2n(4\alpha)^{2/n}N=2\lambda^2n(4\alpha)^{2/n}\left(N+\frac{n}{4}\right).
    \label{oajbjregbiw}
\end{equation}

\subsection{Inflationary Observables\label{aosdbfaisbdfiasdf}}

We can constrain the parameters of our theory by imposing the observational data obtained by Plank \citep{Aghanim:2018eyx}. We list here the experimental values that we use. The amplitude of the dimensionless power spectrum of the scalar perturbations is 
\begin{eqnarray}
    A_s=(2.096\pm0.101)\times 10^{-9}.
    \label{aoergniujnqweed}
\end{eqnarray}

Its tilt at $1\sigma$ is
\begin{eqnarray}
    n_s=0.9661\pm 0.0040,
    \label{aoidfbiadf}
\end{eqnarray}
while at $2\sigma$ is 
\begin{eqnarray}
    n_s=0.9645\pm 0.0096.
    \label{aodgbuawefde}
\end{eqnarray}

\begin{figure}[b]
\includegraphics[width =85mm]{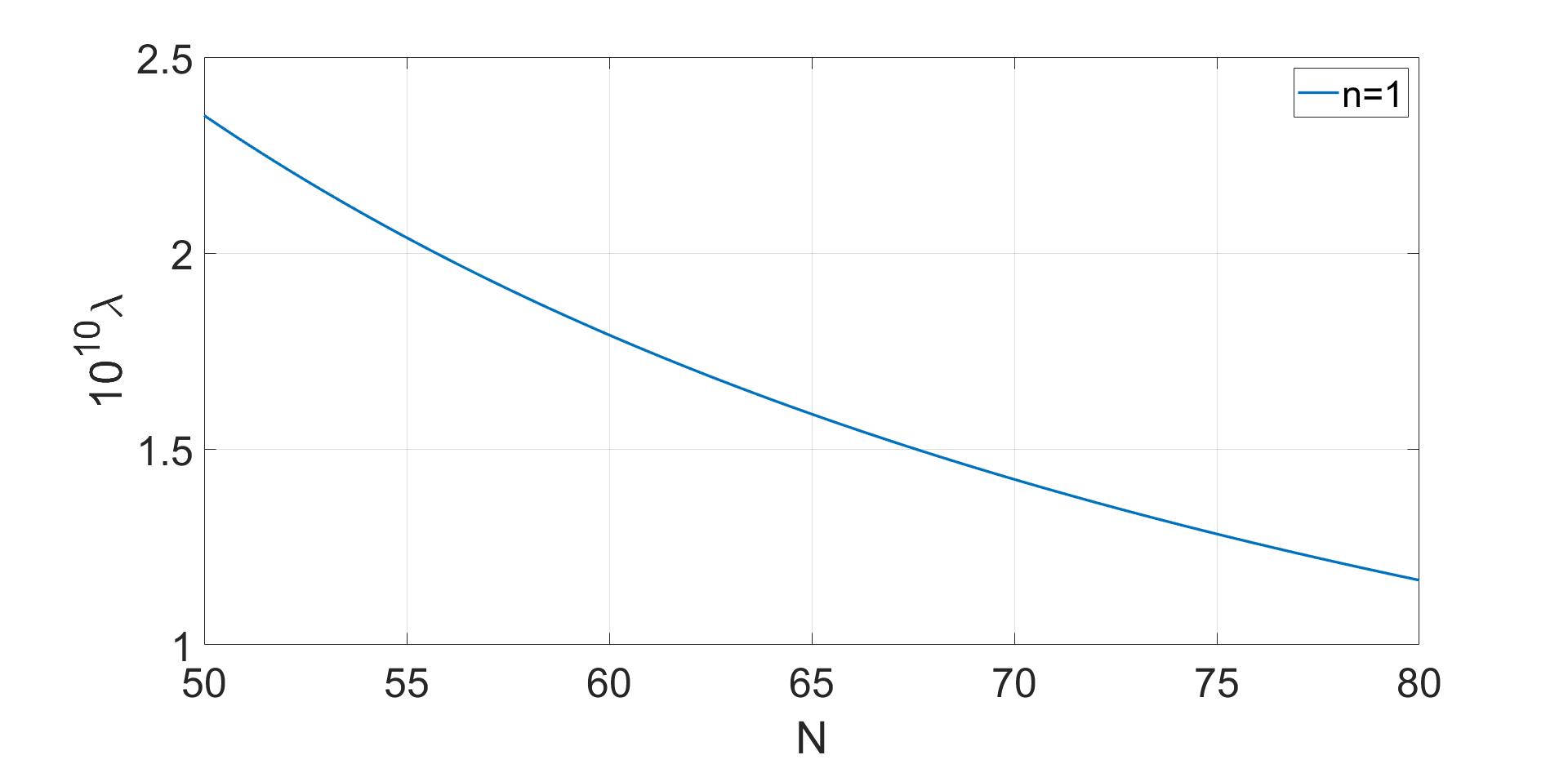}
\caption{\label{asdofnansdfsbsbs} Constant $\lambda^n$ for $n=1$ as a function of the number of e-folds $N$ in the range of interest for quitenssential inflation.}
\end{figure}

\begin{figure}[b]
\includegraphics[width =85mm]{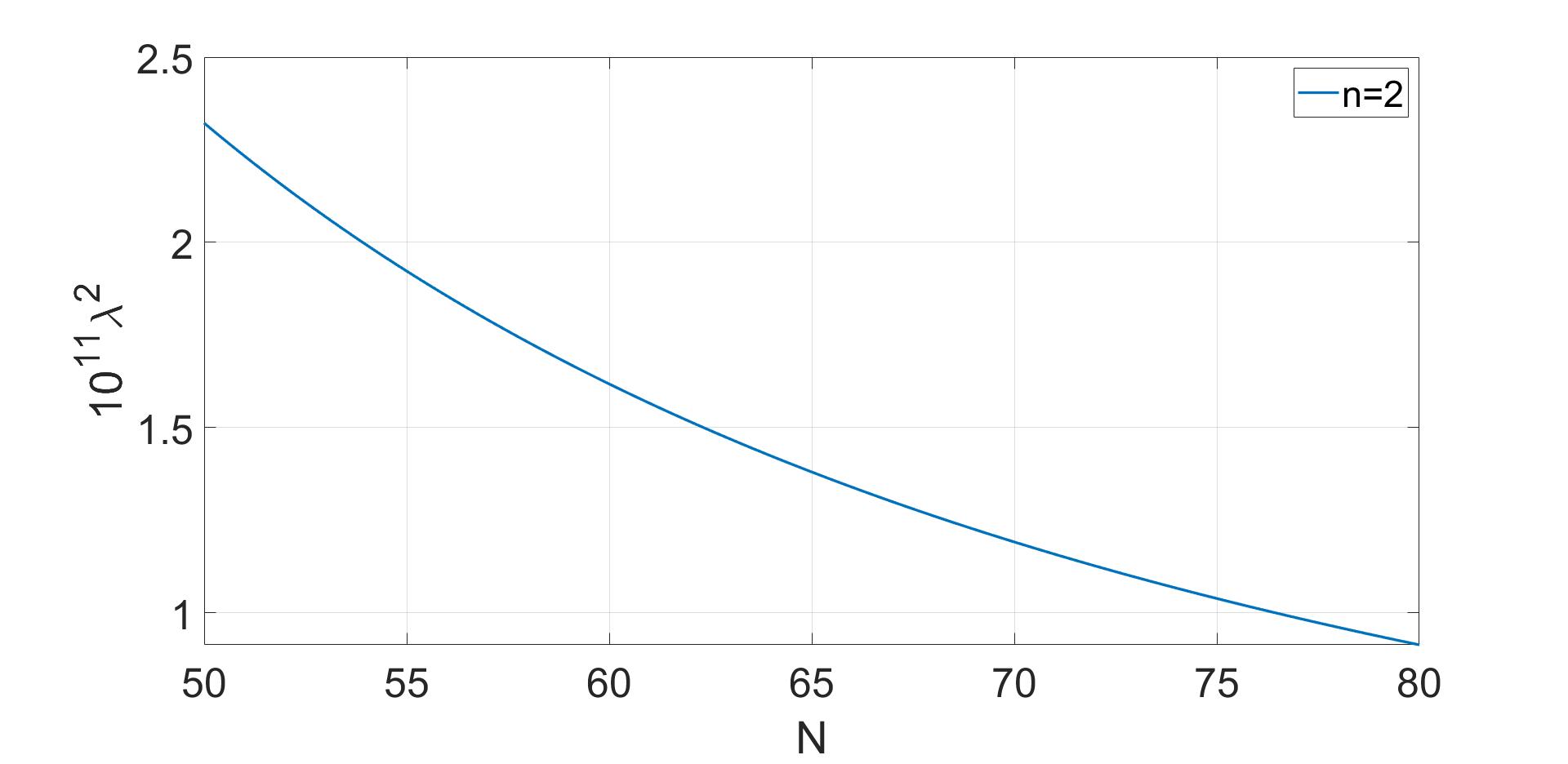}
\caption{\label{apihdhbfhuabsdfasdf} Constant $\lambda^n$ for $n=2$ as a function of the number of e-folds $N$ in the range of interest for quitenssential inflation.}
\end{figure}

\begin{figure}[b]
\includegraphics[width =85mm]{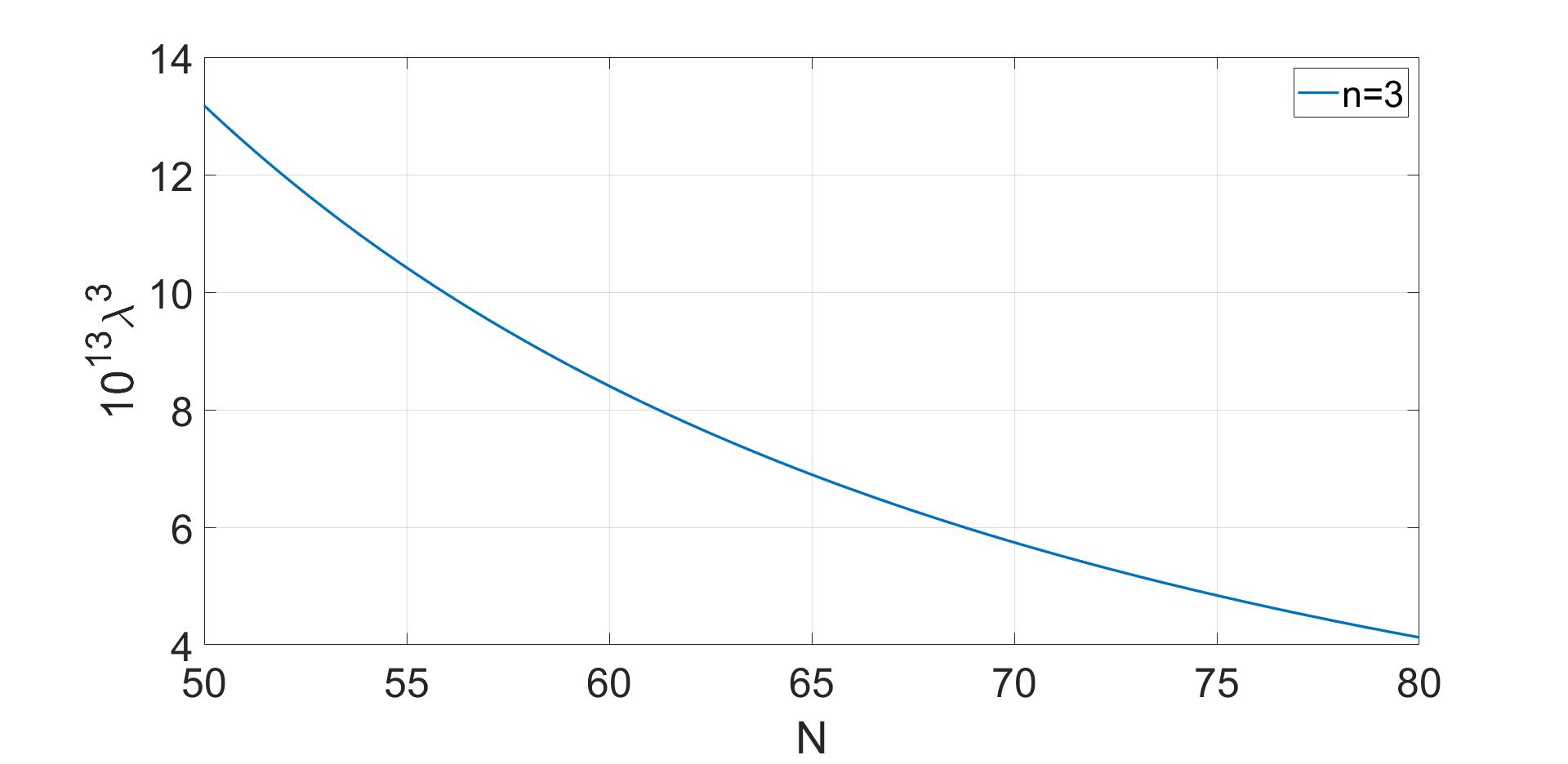}
\caption{\label{ajgjbahbdfawdf} Constant $\lambda^n$ for $n=3$ as a function of the number of e-folds $N$ in the range of interest for quitenssential inflation.}
\end{figure}

\begin{figure}[b]
\includegraphics[width =85mm]{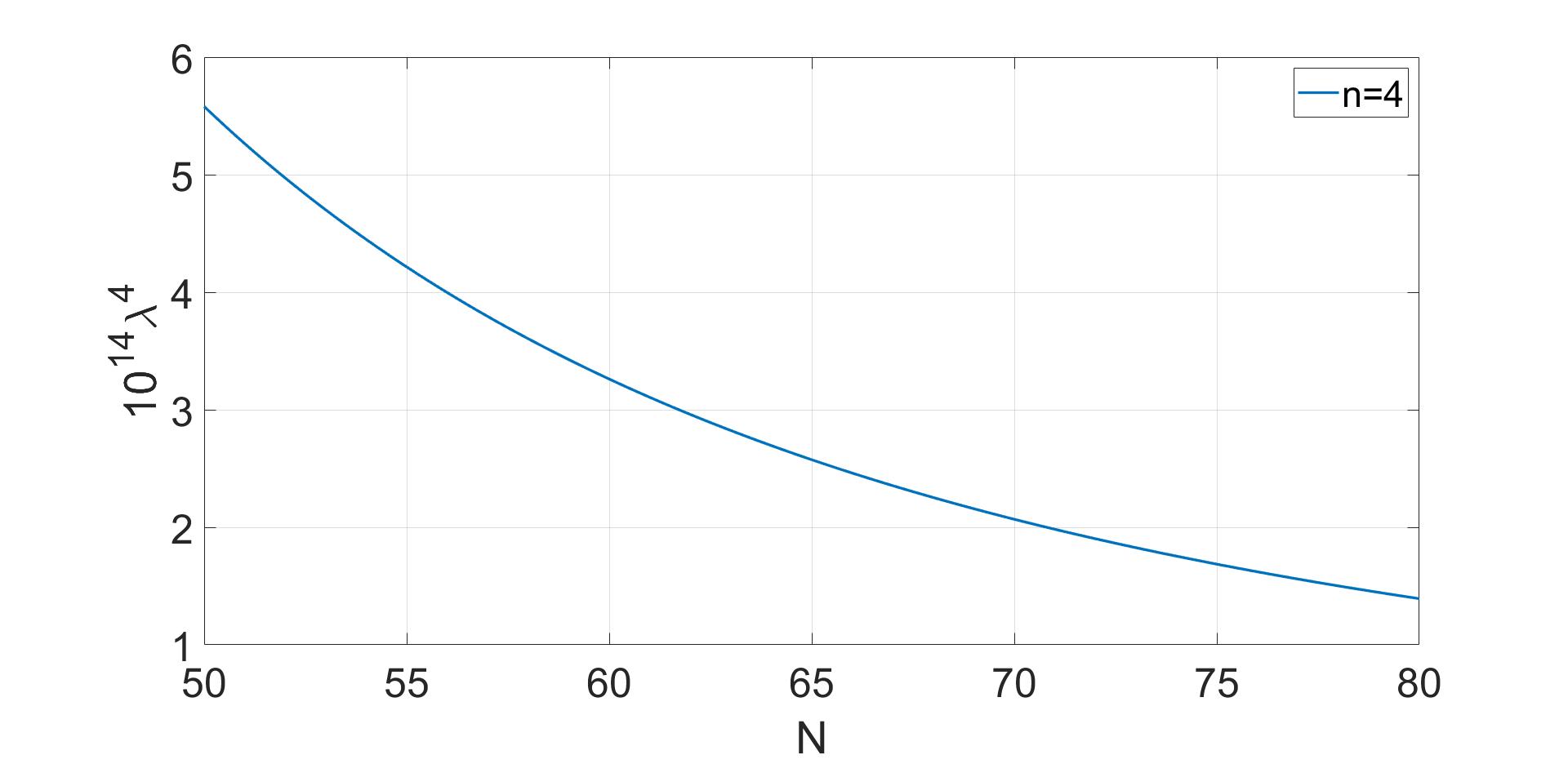}
\caption{\label{aajewfaibwdfhauwefb} Constant $\lambda^n$ for $n=4$ as a function of the number of e-folds $N$ in the range of interest for quitenssential inflation.}
\end{figure}

Furthermore, the tensor-to-scalar ratio is constrained to be
\begin{eqnarray}
    r=\frac{A_h}{A_s}<0.056.
\end{eqnarray}

Let us start with the curvature power spectrum. In the slow-roll approximation, which is valid at the time at which the cosmological scales exit the horizon, it reads
\begin{eqnarray}
    &&A_s=\frac{\bar{V}(\phi_{*})}{24\pi^2\M^4\epsilon_V(\phi_{*})}=\frac{x_{*}^{n+2}}{48\pi^2n^2\lambda^2\alpha(4\alpha)^{2/n}}\Rightarrow\nonumber\\
    &&A_s=\frac{2^{n/2}n^{n/2-1}}{6\pi^2}\lambda^n\left(N+\frac{n}{4}\right)^{\frac{n+2}{2}},
    \label{aibgiWE}
\end{eqnarray}
where we have used the first Friedmann equation and Eq. \eqref{oajbjregbiw} for the value of the field $x$ at horizon exit. It follows that $A_s$ is independent of $\alpha$. Note that the total number of e-folds $N$ depends on the specific details of the kination period. See Figs. \ref{asdofnansdfsbsbs}, \ref{apihdhbfhuabsdfasdf}, \ref{ajgjbahbdfawdf} and \ref{aajewfaibwdfhauwefb} for graphs representing the constant $\lambda^n$ for different values of $n$ as a function of the number of e-folds $N$. Note that in quintessential inflation, we typically have $N\in[60,70]$.

The scalar spectral index reads
\begin{eqnarray}
    n_s&&=1-6\epsilon_V(\phi_{*})+2\eta_V(\phi_{*})=\nonumber\\
    &&=1-\lambda^2(4\alpha)^{2/n}\frac{n(n+2)+n(n+2)x_{*}^n}{x^2_{*}(1+x^n_{*})}\nonumber\\
    &&=1-\lambda^2(4\alpha)^{2/n}\frac{n(n+2)}{x_{*}^2}\Rightarrow\nonumber\\
    &&n_s=1-\frac{n+2}{2\left(N+\frac{n}{4}\right)},
    \label{IARBGUHABRG}
\end{eqnarray}
where we have used Eqs. \eqref{oiabwergib}, \eqref{aijrbghubaweeee} and \eqref{oajbjregbiw}. It follows that the scalar spectral index depends only on the number of e-folds (and on $n$) and does not depend on the parameters of the theory $\alpha$, $M$ and $\lambda$. Remember the remaining number of inflationary e-folds $N$ depends on the details of the kination period.

Finally, it is straightforward to obtain that the tensor-to-scalar ratio reads
\begin{eqnarray}
    r&&=16\epsilon_V(\phi_{*})=\lambda^2(4\alpha)^{2/n}n^2\frac{8}{x_{*}^2(1+x_{*}^n)}\nonumber\\
    &&=\frac{4n}{\left(N+\frac{n}{4}\right)\left(1+x_{*}^n\right)}\Rightarrow\nonumber\\
    &&r=\frac{4n}{\left(N+\frac{n}{4}\right)}\frac{1}{\left[1+4(2n)^{n/2}\lambda^n\alpha\left(N+\frac{n}{4}\right)^{n/2}\right]},
    \label{aigbaerg}
\end{eqnarray}
where we have used Eq. \eqref{oajbjregbiw}.

\begin{figure}[b]
\includegraphics[width =85mm]{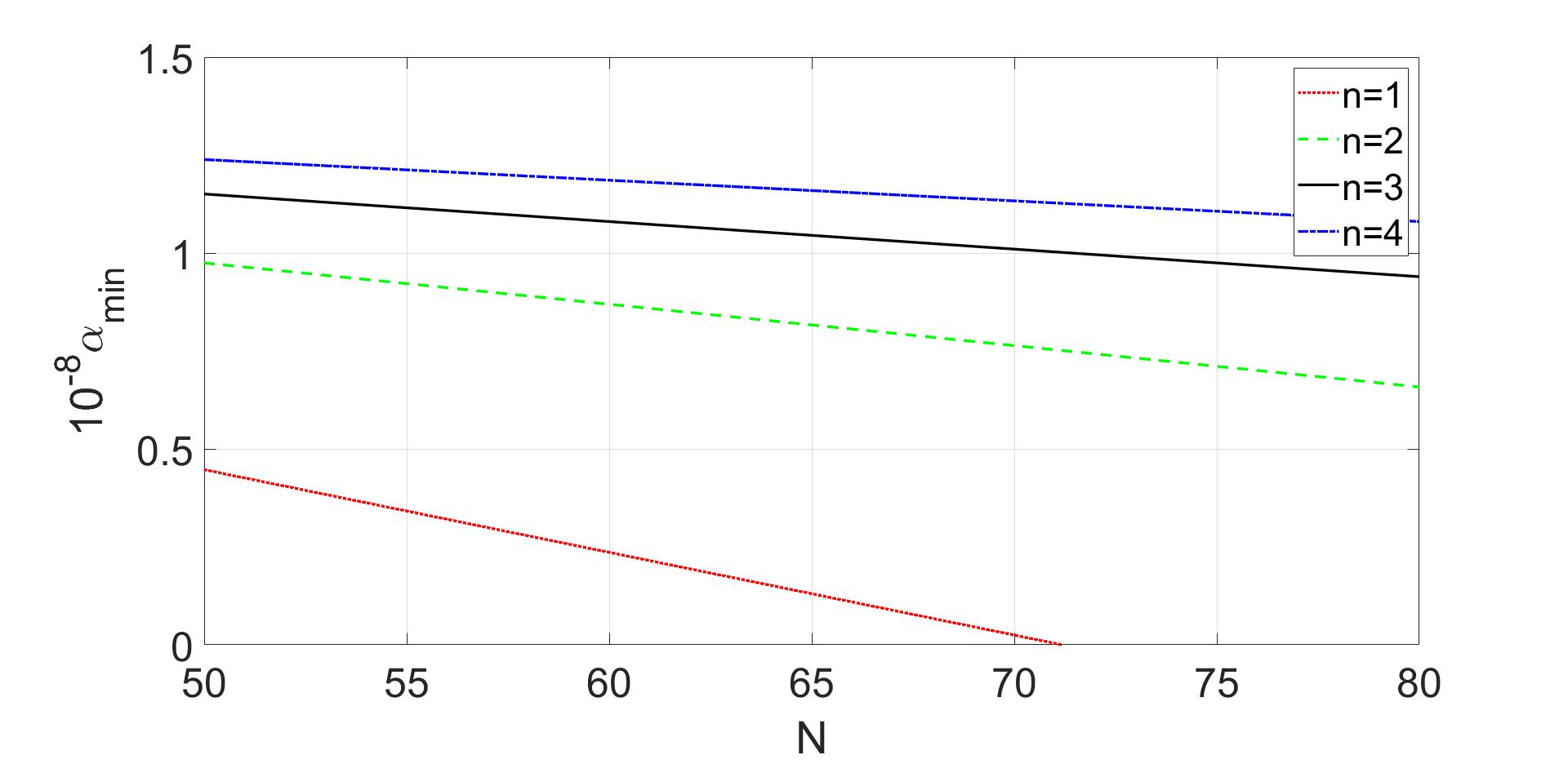}
  \caption{\label{ajbahuiwbefawef} Lower bound on $\alpha$ as a function of the number of e-folds $N$ for $n=1$ (red dotted line), $n=2$ (green dashed line), $n=3$ (black solid line) and $n=4$ (blue dash-dot line) obtained by imposing $r=0.056$. The lower bound is roughly $\alpha\sim 10^8$ for all values of $n$
for the typical number of e-folds in quintessential inflation models $N\in[60,70]$.}
\end{figure}

\begin{figure}[b]
\includegraphics[width =85mm]{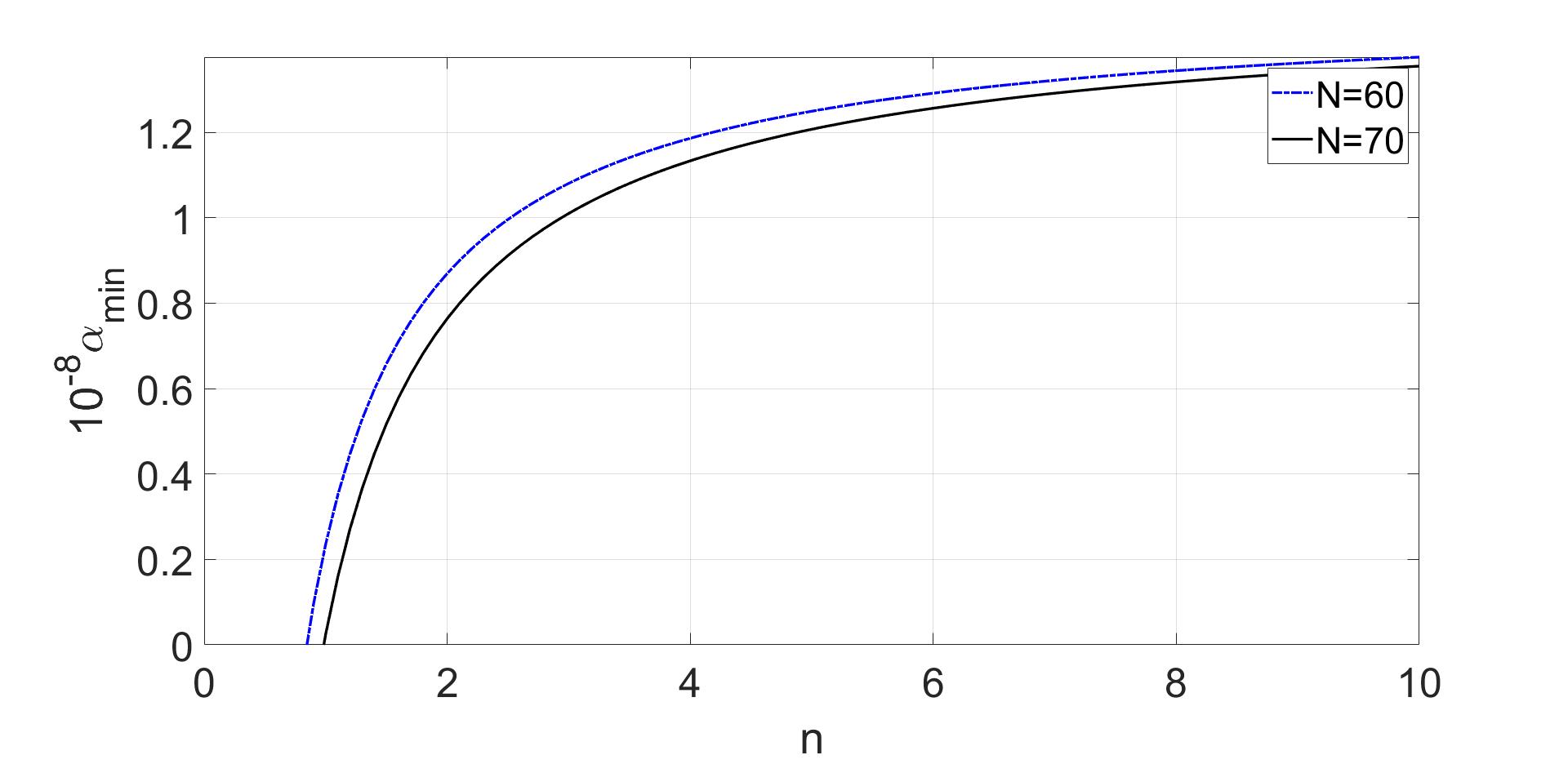}
\caption{\label{aushdbhfuoawbefa} Lower bound on $\alpha$ as a function of $n$ for $N=60$ (blue dash-dot line) and $N=70$ (black solid line), obtained by imposing $r=0.056$. The bound quickly becomes insensitive to the specific value of $n$ taken, independently of the number of e-folds within the range of interest in quintessential inflation.}
\end{figure}

To better understand the role $\alpha$ plays in the obesrvational bound $r<0.056$, one can solve for $\lambda^n$ in Eq. \eqref{aibgiWE} and plug it in Eq. \eqref{aigbaerg} to obtain
\begin{eqnarray}
    r=\frac{16n}{(4N+n)}\frac{1}{\left[1+\frac{96\pi^2n}{(4N+n)}\alpha A_s\right]}.
    \label{oasdfiansdfasdf}
\end{eqnarray}

Therefore, $\alpha$ in terms of $r$ reads
\begin{eqnarray}
    \alpha=\frac{\frac{16}{r}-\left(1+\frac{4N}{n}\right)}{96\pi^2 A_s}.
    \label{aogbanweefd}
\end{eqnarray}

This means that $\alpha$ can be small (of order unity) when 
\begin{eqnarray}
  r\simeq \frac{16}{1+\frac{4N}{n}}.
  \label{cancell}
\end{eqnarray}

For $n=2$ and taking taking into account that the existence of a kination period means that the total number of e-folds is typically within the interval $N\in [60,70]$, $\alpha$ is small (of order unity) when the scalar-to-tensor ratio is approximately in the interval  
\begin{eqnarray}
    r\in [0.113,0.132].
\end{eqnarray}

The $1\sigma$ bound $r<0.056$ does not allow this, but it might be marginally allowed at $2\sigma$, where $r<0.114$\citep{Aghanim:2018eyx}.

For $n=1$ accompanied by a long period of kination such that $N\simeq 71$, we have
\begin{eqnarray}
    r=0.055,
\end{eqnarray}
which is marginally within the $1\sigma$ bounds. See Fig. \ref{oareygboaebrf} for the $r-n_s$ graph in the $n=1$ case. Note however, that we expect \mbox{$N\lesssim 70$} or so, for otherwise kination lasts too long and there is danger that a spike in the spectrum of primordial gravitational waves, corresponding to the scales which reenter the horizon during kination, threatens to destabilise Big Bang Nucleosynthesis \citep{Agarwal:2017wxo}.

When the tensor-to-scalar ratio takes the value given by Eq.~\eqref{cancell}, $\alpha$ can be very small (of order unity). However, as we have explained above, this in general not the case (when $n=\mathcal{O}(1)$ we have $\alpha\gtrsim 10^8$ as can be seen in Figs. \ref{ajbahuiwbefawef} and \ref{aushdbhfuoawbefa})). Indeed, using $N=60$ and $A_s=2\times10^{-9}$, we have the following bounds for some values of $n$ by imposing $r<0.056$
\begin{eqnarray}
        &&n=2\quad \Rightarrow\quad \alpha>0.87\times10^8\nonumber\\
        &&n=4\quad \Rightarrow\quad \alpha>1.18\times10^8\nonumber\\
        &&n=8\quad \Rightarrow\quad \alpha>1.34\times10^8.
        \label{aojdjgbiawer}
\end{eqnarray}
Note that $\alpha$ is a non-perturbative coefficient that can be much larger than unity without a problem. Note also that these bounds are a direct consequence of the observational value of the scalar power spectrum and cannot be realaxed via the choice of a suitable value of~$\lambda^n$.

We end this section with a remark regarding the Lyth bound \citep{lythbound}. By expressing the equation of motion of the inflaton (during slow-roll) as a function of the number of e-folds, with the help of Eq. \eqref{oasdfiansdfasdf}, it is straightforward to obtain the variation of the inflaton from the time at which the cosmological scales exit the horizon until the end of inflation as
\begin{eqnarray}
    \Delta \phi&&\equiv \phi_{\text{end}}-\phi_{*}\nonumber\\
    &&=-\M\sqrt{2n}\int_{N_{*}}^{N_{\text{end}}}\frac{\text{d}N}{\sqrt{4N+n+96\pi^2n\alpha A_s}}.
    \label{asdfeeawerawdf}
\end{eqnarray}

We consider two different limits. Firstly, for $\alpha$ at least one order of magnitude larger than $A_s^{-1}$, \textit{e.g.}, the higher bound $\alpha=10^{10}$ in Figs. \eqref{oareygboaebrf}-\eqref{qpiuerbgbhubqaedf}, $r$ is very small (cf. Eq. \eqref{oasdfiansdfasdf}) and the third term in the square root in Eq. \eqref{asdfeeawerawdf} dominates. It can then be easily found that the displacement of the inflaton, taking $N_{*}-N_{\text{end}}\simeq 70$, as is usual in quintessential inflation, reads
\begin{equation}
    \Delta \phi \sim 0.7\M. 
\end{equation}
Note that for arbitrarily large $\alpha$, $\Delta \phi$ can be made arbitrarily small, \textit{e.g.}, for $\alpha \sim 10^{13}$ we have $\Delta \phi \sim 10^{-2}\M$.

In the opposite limit, when the value of $\alpha$ is around the lower bound given by Eq. \eqref{aojdjgbiawer}, all terms in the square root in Eq. \eqref{asdfeeawerawdf} are comparable. However, the integration can easily be carried out, yielding, for $n\sim \mathcal{O}(1)$ and $N=70$,
\begin{equation}
    \Delta \phi \sim 6 \M.
    \label{adsofiabnsdfasdf}
\end{equation}

In order to obtain the displacement of the canonical field in the Jordan frame $\varphi$, for a given $n$, we would need to integrate Eq. \eqref{aourubgiuawef} to obtain the relation between $\varphi$ and $\phi$. In general it is not possible to obtain an analytic expression, except for the $n=2$ case. This case is studied in detail below and the displacement of $\varphi$ is calculated there.

\section{Kination} \label{aieurgbawe}

\subsection{Dynamics in the Jordan and Einstein Frames}

After the inflaton reaches the value given by Eq. \eqref{aogiubwefd} and inflation ends, a new cosmological era called kination starts. During kination, the dominant contribution to the energy density of the Universe is still that of the inflaton. Furthermore, as the slope the potential becomes larger in magnitude, the inflaton becomes oblivious to the potential and its energy density is dominated by the kinetic part. Varying the action \eqref{aijbgaweref} with respect to $\varphi$ we obtain the usual Klein-Gordon equation (remember in the Jordan frame the field is minimally coupled to gravity). Thus, during kination, the equation of motion of the inflaton reads
\begin{eqnarray}
    \Ddot{\varphi}+3H\Dot{\varphi}\simeq 0.
    \label{aoaubewfwef}
\end{eqnarray}

This equation can be readily integrated to obtain 
\begin{eqnarray}
    \Dot{\varphi}\propto a^{-3} \Leftrightarrow \rho_{\varphi}=\frac{1}{2}\Dot{\varphi}^2 \propto a^{-6}.
    \label{agruauwef}
\end{eqnarray}

However, although the dominant contribution to the energy density of Universe is that of the inflaton, Eq. \eqref{agruauwef} in general is not the energy density of the Universe. This is because, in the Palatini formalism, new effective matter sources are introduced as a consequence of the Starobinski term in the action. We can see this by calculating the zeroth-zeroth component of the Einstein equations \eqref{aofjgbaiwner}. Using Eq. \eqref{aorugbaibwef} and remembering that during a kinetic dominated era the kinetic energy density of the inflaton is bounded as\citep{Meng:2004yf}
    \begin{eqnarray}
        \frac{1}{2}\Dot{\varphi}^2<\frac{\M^4}{2\alpha},
        \label{aodfbausdf}
    \end{eqnarray}
which means that 
\begin{eqnarray}
    f_R^{-1}(R)=\left(1-\frac{\alpha}{\M^4}\Dot{\varphi}^2\right)^{-1}\simeq 1+\frac{\alpha}{\M^4}\Dot{\varphi}^2,
\end{eqnarray}
the zeroth-zeroth component of the Einstein equations reads
\begin{eqnarray}
    3H^2&&=\frac{\Dot{\varphi}^2}{2\M^2}+\frac{6H\alpha}{\M^4}\dot{\varphi}\ddot{\varphi}+\frac{3\alpha}{4\M^6}\dot{\varphi}^4+\frac{3\alpha^2}{\M^8}\dot{\varphi}^2\ddot{\varphi}(2H\dot{\varphi}-\ddot{\varphi})\nonumber\\
    &&-\frac{\alpha^2}{4\M^{10}}\dot{\varphi}^6 -\frac{6\alpha^3}{\M^{12}}\dot{\varphi}^4\ddot{\varphi}^2.
\end{eqnarray}

This equation can be further simplified by using Eq.~\eqref{aoaubewfwef} to obtain 
\begin{eqnarray}
    3H^2\M^2&=&\frac12{\dot{\varphi}^2}-\frac{2\alpha}{\M^2}\ddot{\varphi}^2+\frac{3\alpha}{4\M^4}\dot{\varphi}^4-\frac{5\alpha^2}{\M^6}\dot{\varphi}^2\ddot{\varphi}^2\nonumber\\
    &&-\frac{\alpha^2}{4\M^8}\dot{\varphi}^6-\frac{6\alpha^3}{\M^{10}}\dot{\varphi}^4\ddot{\varphi}^2\nonumber\\
   &=&       \frac12\dot\varphi^2\left[1+\frac{3\alpha}{2m_{\rm P}^4}\dot\varphi^2
      \left(1-\frac{\alpha}{3m_{\rm P}^4}\dot\varphi^2\right)\right]\nonumber\\
    &&       -\frac{2\alpha}{m_{\rm P}^2}\ddot\varphi^2\left[1+
      \frac{5\alpha}{2m_{\rm P}^4}\dot\varphi^2
      \left(1+\frac{6\alpha}{5m_{\rm P}^4}\dot\varphi^2\right)\right]
\end{eqnarray}
  If $\alpha$ is not very large, Eq.~\eqref{aodfbausdf} can be strongly satisfied, especially as the kinetic energy density decreases rapidly after the end of inflation, cf. Eq.~\eqref{agruauwef}. Then, the above is reduced to
\begin{eqnarray}
      3H^2m_{\rm P}^2&\simeq&
              \frac12\dot\varphi^2-\frac{2\alpha}{m_{\rm P}^2}\ddot\varphi^2
  \nonumber\\
  &\simeq&
  \frac12\dot\varphi^2\left(1-36\alpha\frac{H^2}{m_{\rm P}^2}\right),
  \label{HH}
\end{eqnarray}
where we also used Eq.~\eqref{aoaubewfwef}\footnote{Rearranging Eq.~\eqref{HH}
  we obtain \mbox{$(\dot\varphi/H)^2=6\M^2+36\alpha\dot\varphi^2/\M^2$}, where
  one can see that $H$ is not zero for any finite value of $\alpha$, that is
the brackets in Eq.~\eqref{HH} are always positive.}.
$H$ is diminishing with time, so $H^2<H_{\rm inf}^2\sim 10^{-10}m_{\rm P}^2$. Thus, if $\alpha$ is not too large, the second term in the parenthesis above very soon becomes negligible compared to unity. It follows that the main contribution to the energy density of the Universe is the kinetic energy density of the inflaton
\begin{eqnarray}
    3H^2\M^2\simeq \frac{1}{2}\dot{\varphi}^2.
\end{eqnarray}

Using Eq. \eqref{agruauwef}, we have
\begin{equation}
    \rho = \rho_{\varphi}=\frac{1}{2}\dot{\varphi}^2\propto a^{-6} \Leftrightarrow w=1 \Leftrightarrow a\propto t^{1/3}\Leftrightarrow H=\frac{1}{3t},
    \label{aoubbiawedf}
\end{equation}
where $w$ is the barotropic parameter of the Universe.

We conclude that the modifications to the kination dynamics coming from the introduction of a Starobinski term in Palatini $f(R)$ gravity are subdominant and the typical situation is recovered. 

Equivalent conclusions can be obtained in the Einstein frame. Indeed, close to the origin, the modified Peebles-Vilenkin potential reads 
\begin{eqnarray}
    V(\varphi)\simeq \frac{\lambda^n M^n}{\M^{n-4}},
    \label{aodbabhidf}
\end{eqnarray}
so that the field redefinition \eqref{aodbabiwdf} for the (non-canonical) kinetic term in the action \eqref{aoiodfgjbairg} now reads 
\begin{eqnarray}
    \text{d}\phi=\frac{\text{d}\varphi}{\sqrt{1+\frac{4\alpha\lambda^n M^n}{\M^n}}}\simeq \left(1-\frac{2\alpha\lambda^n M^n}{\M^n}\right)\text{d}\varphi,
\end{eqnarray}
where we have used that $M\ll \M$ and $\alpha \lambda^n\ll1 $ (see below). It follows that the kinetic term of $\varphi$ is canonical to a very good approximation, \textit{i.e.}, $\phi\simeq\varphi$. Furthermore, the coupling in the matter action does not affect the dynamics (see the discussion after Eq. \eqref{adofbbasdf}). Thus, since the inflaton is still oblivious to the potential, in the Einstein frame we have the equation
\begin{eqnarray}
    \ddot{\phi}+3\bar{H}\dot{\phi}\simeq 0.
\end{eqnarray}

As for the zeroth-zeroth component of the Einstein equations, from Eq. \eqref{aoijdbfaiusdf} we have
\begin{eqnarray}
    3\bar{H}^2\M^2=\frac{1}{2}\dot{\phi}^2+\frac{3\alpha \dot{\phi}^4}{4\M^4}+\frac{\alpha^2\dot{\phi}^6}{2\M^8}.
    \label{adbabbidfa}
\end{eqnarray}
where barred quantities are calculated using the metric in the Einstein frame \eqref{aiffgjbaierg} and dots represent $\text{d}/\text{d}\bar{t}$.

Again, using Eq. \eqref{aodfbausdf} and $\phi\simeq \varphi$ during kination, the Friedmann equation reads, to a very good approximation,
\begin{eqnarray}
    3\bar{H}^2\M^2 \simeq \frac{1}{2}\dot{\phi}^2.
\end{eqnarray}

\subsection{Reheating and Number of e-folds }

When there is a cosmological era after inflation with a stiff equation of state with barotropic parameter $w$, the number of inflationary e-folds is increased by \citep{Dimopoulos:2019gpz}
\begin{eqnarray}
    \Delta N=\frac{3w-1}{3(1-w)}\ln{\left(\frac{V_{\text{end}}^{1/4}}{T_{\text{reh}}}\right)}.
    \label{awdoufawef}
\end{eqnarray}

In common inflationary models, after inflation ends, the Universe is perturbately reheated when the inflaton oscillates around the minimum of its potential. It is easy to show that in this situation the effective barotropic of the Universe is $w=0$, so that the prefactor in Eq. \eqref{awdoufawef} is $-1/3$ and the remaining e-folds of inflation are actually decreased. In contrast, during kination, the barotropic parameter of the Universe is $w=1$ (see Eq.~\eqref{aoubbiawedf}), so that the prefactor is $+1/3$. Thus, the remaining number of inflationary e-folds is increased by
\begin{eqnarray}
    \Delta N=\frac{1}{3}\ln{\left(\frac{\bar{V}^{1/4}(\phi_{\text{end}})}{T_{\text{reh}}}\right)},
\end{eqnarray}
where $T_{\text{reh}}$ is the temperature of the radiation bath at reheating and $\bar{V}(\phi_{\text{end}})$ is the potential at the end of inflation. In this way, in what follows we consider that the remaining number of inflationary e-folds after the cosmological scales exit the horizon is given by
\begin{eqnarray}
    N=60+\Delta N.
\end{eqnarray}

The lowest value for $T_{\text{reh}}$, and, therefore, the highest for $\Delta N$, is obtained through gravitational reheating (for which reheating occurs at the end of inflation $t_{\text{reh}}=t_{\text{end}}$)\footnote{It is important to mention that modifications to the gravitational particle production, due to the $R^2$ term in the action, are possible. However, during inflation this term and the Einstein-Hilbert one are comparable. Therefore, any possible modifications are of order unity. This is why, for simplicity, we assume the dominant contribution comes from the latter. The study of particle production due to an event horizon in Palatini $f(R)$ gravity will be addressed in a future work.}. For this reheating mechanism, it can be shown \citep{Dimopoulos:2017zvq} that
\begin{eqnarray}
    T_{\text{reh}}^{\text{gr}}\sim 10^{-2}\frac{H^2(\phi_{\text{end}})}{\M}.
\end{eqnarray}

Assuming that the slow-roll approximation is still valid at the end of inflation, we have
\begin{eqnarray}
    T_{\text{reh}}^{\text{gr}}=10^{-2}\frac{\bar{V}(\phi_{\text{end}})}{3\M^3}.
\end{eqnarray}

Thus, the increase in the number of e-folds reads
\begin{eqnarray}
    \Delta N&&=\frac{1}{3}\ln{\left(\frac{3\M^3\bar{V}^{1/4}(\phi_{\text{end}})}{10^{-2}\bar{V}(\phi_{\text{end}})}\right)}\nonumber\\
    &&\simeq 2 +
          \ln{\left(\frac{\M}{\bar{V}^{1/4}(\phi_{\text{end}})}\right)}.
\end{eqnarray}

The potential at the end of inflation $\bar{V}(\phi_{\text{end}})$ can be obtained by evaluating Eq. \eqref{aruguauiwebn} at $x_{\text{end}}$, given by Eq. \eqref{aogiubwefd}. It reads 
\begin{equation}
    \bar{V}(\phi_{\text{end}})=\frac{\M^4}{4\alpha}\frac{x^n(\phi_{\text{end}})}{1+x^n(\phi_{\text{end}})}=\frac{\M^4 n^n \lambda^n}{2^{n/2}+4\alpha n^n\lambda^n},
    \label{aifugfbaniwef}
\end{equation}
and the remaining number of e-folds is increased by
\begin{eqnarray}
    \Delta N=2+\frac{1}{4}\ln{\left(\frac{2^{n/2}+4\alpha n^n\lambda^n}{n^n \lambda^n}\right)}.
    \label{oabergiaer}
\end{eqnarray}

Note that, by virtue of Eq. \eqref{adofasiudfhasdf}, Eq. \eqref{aifugfbaniwef} is simplified as
\begin{eqnarray}
    \bar{V}(\phi_{\text{end}})=\frac{\M^4n^n\lambda^n}{2^{n/2}},
    \label{a9udsbfasdf}
\end{eqnarray}
so that Eq. \eqref{oabergiaer} is simplified as
\begin{eqnarray}
  \Delta N=2+
    \frac{n}{4}\ln{\left(\frac{\sqrt 2}{n\lambda}\right)}.
    \label{adfoabsdfasdf}
\end{eqnarray}

We emphasize that Eq.~\eqref{adofasiudfhasdf}, and thus the approximated expressions in Eqs.~\eqref{a9udsbfasdf} and \eqref{adfoabsdfasdf}, only hold when we work near the lower bound for $\alpha$ (as we do in the present work).

From Eq. \eqref{IARBGUHABRG}, taking into account that the remaining number of inflationary e-folds is $N=60+\Delta N$ we have
\begin{eqnarray}
    n_s=1-\frac{n+2}{2\left(60+\Delta N+\frac{n}{4}\right)}.
    \label{aierbrgaewfd}
\end{eqnarray}

At this point, in order to obtain analytical results we need to choose specific values for $n$.

\begin{figure}[b]
\includegraphics[width =84mm]{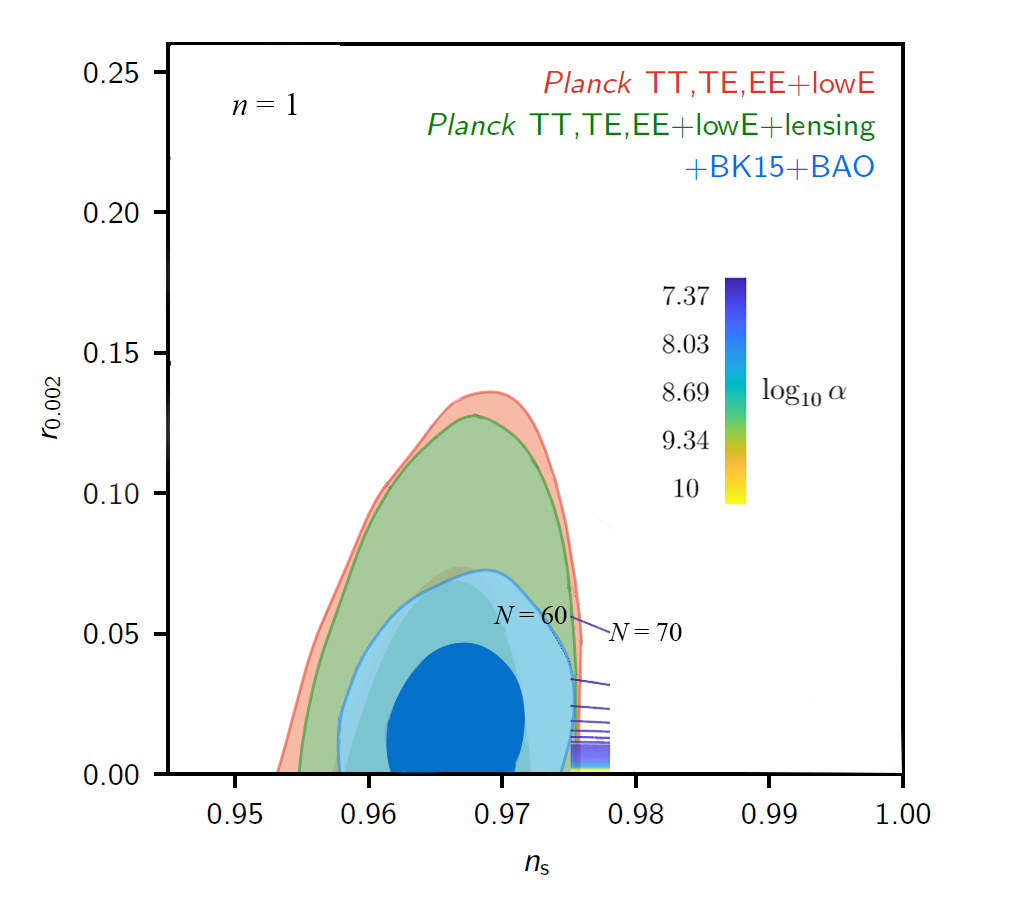}
\caption{\label{oareygboaebrf} $r-n_s$ graph where the predictions derived from our model, for $n=1$, are compared to the experimental data. The number of e-folds represented range from 60 (left side) to 70 (right side). The parameter $\alpha$ ranges from its lower bound $\alpha_{\text{min}}=2.36\times10^7$ (blue) to $\alpha=10^{10}$ (yellow). Figure adapted from Ref.~\citep{Aghanim:2018eyx}.}
\end{figure}

\begin{figure}[b]
\includegraphics[width =85mm]{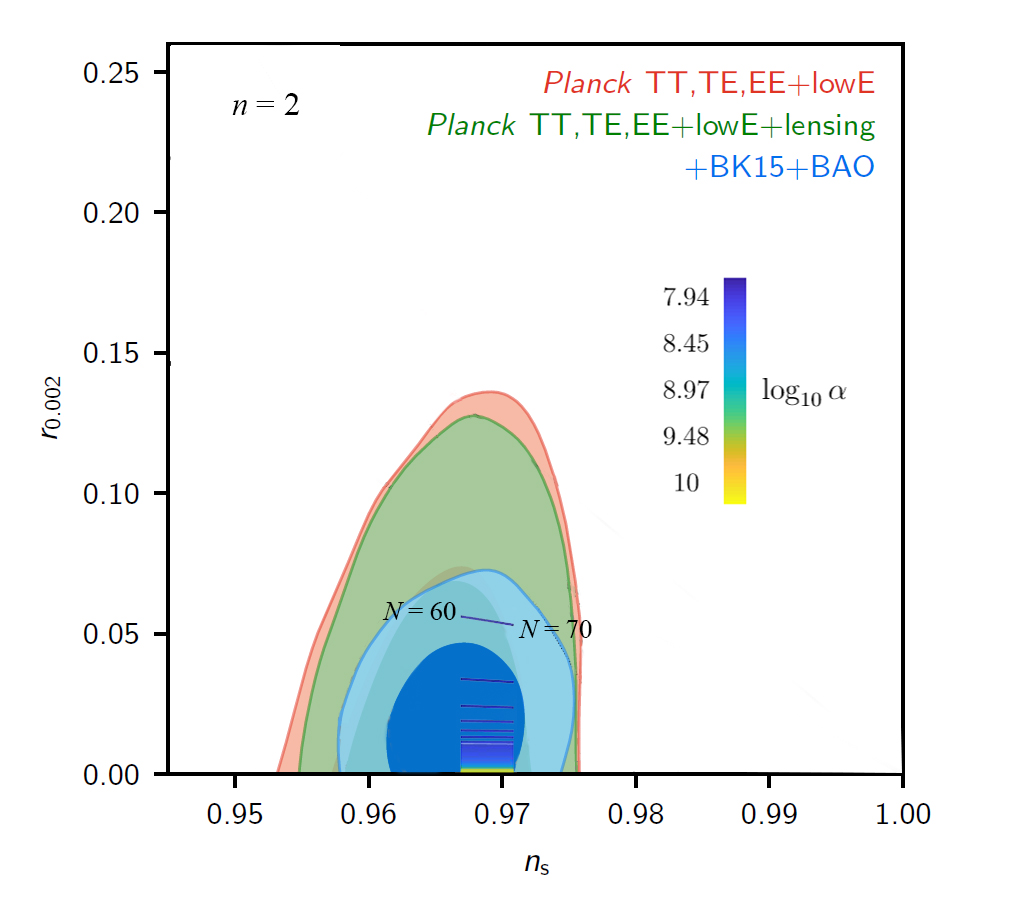}
\caption{\label{qripgbuqerf} $r-n_s$ graph where the predictions derived from our model, for $n=2$, are compared to the experimental data. The number of e-folds represented range from 60 (left side) to 70 (right side). The parameter $\alpha$ ranges from its lower bound $\alpha_{\text{min}}=8.7\times10^7$ (blue) to $\alpha=10^{10}$ (yellow). Figure adapted from Ref.~\citep{Aghanim:2018eyx}.}
\end{figure}

\begin{figure}[b]
\includegraphics[width =85mm]{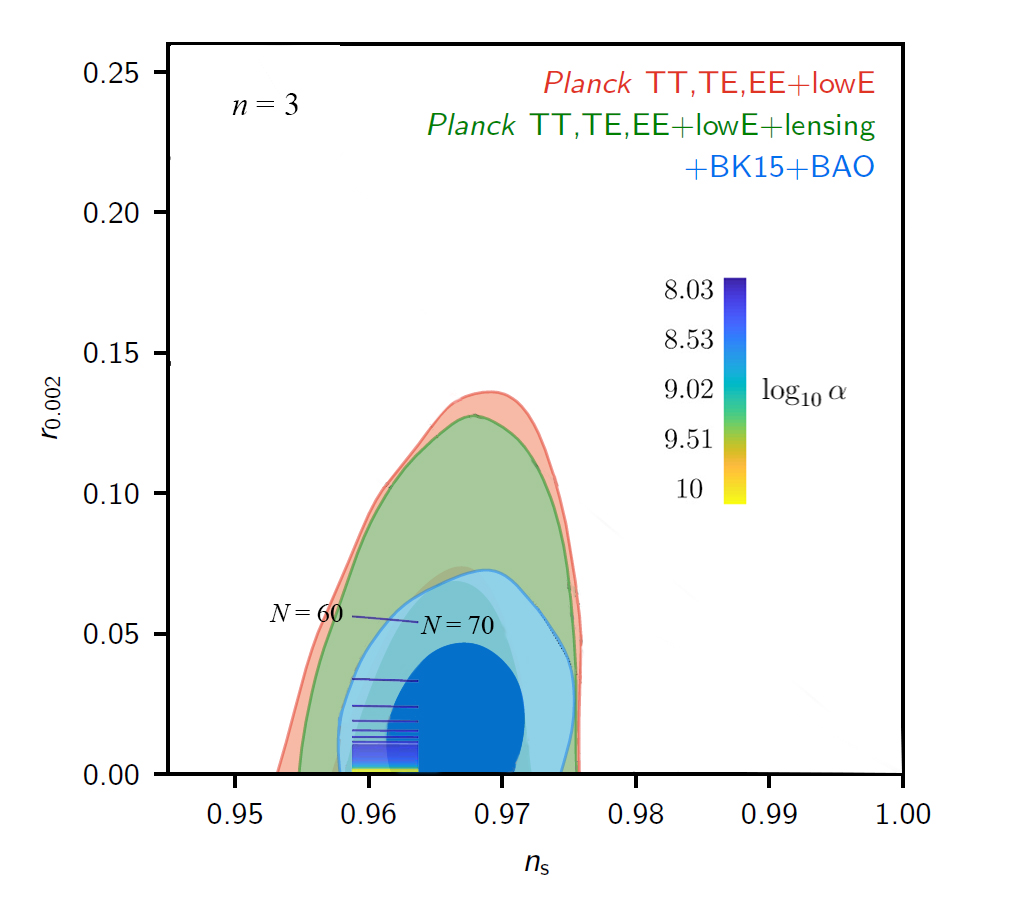}
\caption{\label{airbahuebrfwwww} $r-n_s$ graph where the predictions derived from our model, for $n=3$, are compared to the experimental data. The number of e-folds represented range from 60 (left side) to 70 (right side). The parameter $\alpha$ ranges from its lower bound $\alpha_{\text{min}}=1.08\times10^8$ (blue) to $\alpha=10^{10}$ (yellow). Figure adapted from Ref.~\citep{Aghanim:2018eyx}.}
\end{figure}

\begin{figure}[b]
\includegraphics[width =85mm]{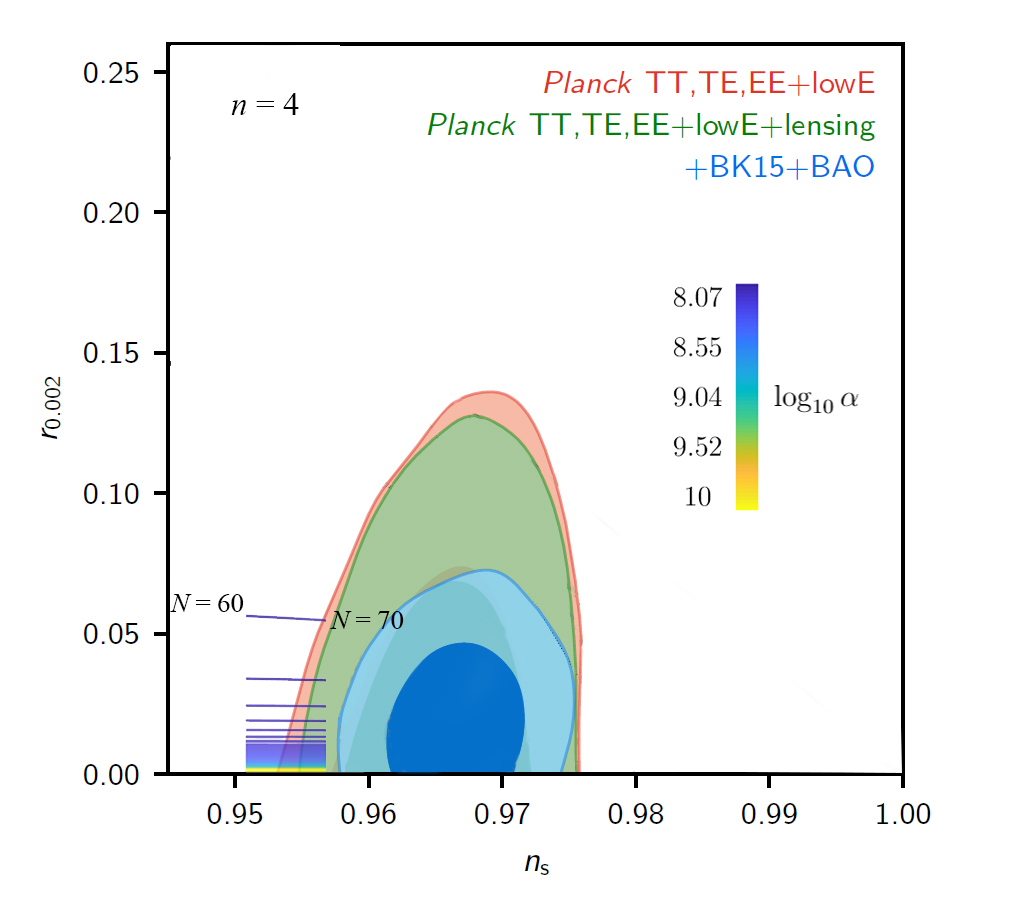}
\caption{\label{qpiuerbgbhubqaedf} $r-n_s$ graph where the predictions derived from our model, for $n=4$, are compared to the experimental data. The number of e-folds represented range from 60 (left side) to 70 (right side). The parameter $\alpha$ ranges from its lower bound $\alpha_{\text{min}}=1.18\times10^8$ (blue) to $\alpha=10^{10}$ (yellow). Figure adapted from  Ref.~\citep{Aghanim:2018eyx}.}
\end{figure}

\subsection{$n=2$}

In this section we focus on the $n=2$ case. The potential in the Jordan frame, remembering $\varphi\gg M$ during inflation, reads 
\begin{eqnarray}
    V(\varphi)=\lambda^2\M^{2}\varphi^2.
\end{eqnarray}

We can redefine the coupling constant as 
\begin{eqnarray}
    \lambda^2\M^2\equiv \frac12{m^2},
\end{eqnarray}
where $m$ is a suitable mass scale. 

It is worth mentioning that for $n=2$ it is possible to obtain an analytical expression for the potential in the Einstein frame. Indeed, the field redefinition \eqref{aourubgiuawef} now reads
\begin{eqnarray}
    \text{d}\phi=\frac{\M}{2\lambda\sqrt{\alpha}}\frac{\text{d}x}{\sqrt{1+x^2}}.
    \label{asodjfaibusdf}
\end{eqnarray}

Integrating this expression we obtain 
\begin{equation}
    \phi(x)=\frac{\M}{2\lambda \sqrt{\alpha}}\sinh^{-1}{x}\Rightarrow x(\phi)=\sinh{\left(\frac{2\lambda \sqrt{\alpha}}{\M}\phi\right)}.
    \label{adofbabsihdf}
\end{equation}

Using this in Eq. \eqref{aruguauiwebn} we obtain the potential in the Einstein frame
\begin{eqnarray}
    \bar{V}(\phi)=\frac{\M^4}{4\alpha}\tanh^2{\left(\frac{2\lambda \sqrt{\alpha}}{\M}\phi\right)}.
\end{eqnarray}

Choosing $n=2$ in Eqs.~\eqref{aibgiWE}-\eqref{aigbaerg}, the inflationary observables now read
\begin{eqnarray}
    A_s=\frac{m^2}{24\pi^2\M^2}(2N+1)^2,
    \label{aogbiwaer}
\end{eqnarray}
\begin{eqnarray}
    n_s=1-\frac{4}{2N+1},
    \label{aourugbbaiubwef}
\end{eqnarray}
and
\begin{eqnarray}
    r=\frac{16}{(2N+1)\left[1+\frac{4m^2\alpha}{\M^2}(2N+1)\right]},
    \label{aoijgbwe}
\end{eqnarray}
where $N=60+\Delta N$ is the total number of inflationary e-folds. Furthermore, Eq. \eqref{aogiubwefd} now reads
\begin{eqnarray}
    x_{\text{end}}^2=4\frac{m^2}{\M^2}\alpha,
\end{eqnarray}
while the increase in the number of e-folds is 
\begin{eqnarray}
    \Delta N=2+\frac{1}{4}\ln{\left(\frac{1+4\alpha \frac{m^2}{\M^2}}{\frac{m^2}{\M^2}}\right)}.
    \label{aierjjghaenrf}
\end{eqnarray}
The above is reduced to \mbox{$\Delta N=2+\frac12\ln\left(m_{\rm P}/m\right)$} when \mbox{$|x_{\rm end}|\ll 1$}.

In order to obtain the most accurate value for $\Delta N$, one can solve for $m^2/\M^2$ in Eq. \eqref{aogbiwaer} and use it in Eq. \eqref{aierjjghaenrf} to obtain the equation
\begin{eqnarray}
    \Delta N=2+\frac{1}{4}\ln{\left[\frac{(121+2\Delta N)^2+96\alpha\pi^2 A_s}{24\pi^2A_s}\right]}.
\end{eqnarray}

Using the lower bound for alpha $\alpha \sim 8.7\times 10^7$, given by Eq.~\eqref{aojdjgbiawer}, and the observational value for the amplitude of the scalar power spectrum, given by Eq.~\eqref{aoergniujnqweed}, this equation can be numerically solved to obtain 
\begin{eqnarray}
    \Delta N=8.103\simeq 8,
    \label{adfibadf}
\end{eqnarray}
which means that the total number of inflationary e-folds is 
\begin{eqnarray}
    N\simeq 68.
    \label{aierhbghuaberf}
\end{eqnarray}

Using this result in Eq.~\eqref{aourugbbaiubwef} immediately gives 
\begin{eqnarray}
    n_s=0.9708,
\end{eqnarray}
which is slightly larger than the upper $1\sigma$ bound in Eq.~\eqref{aoidfbiadf} but could be easily accommodated by the $2\sigma$ bounds  in Eq.~\eqref{aodgbuawefde}. This can be understood as follows. From Eq.~\eqref{aourugbbaiubwef}, the number of e-folds in terms of $n_s$ reads
\begin{eqnarray}
    N=\frac{1}{2}\left(\frac{2}{1-n_s}-1\right),
\end{eqnarray}
so that the $1\sigma$ bounds correspond to 
\begin{eqnarray}
    N\in[52,66].
    \label{adofaibdfawfd}
\end{eqnarray}

Thus, the extra $6$ e-folds at the upper bound could be explained by a period of kination, although $\Delta N=8$ would be too large to be within the $1\sigma$ bounds.

The mass scale $m^2$ is fixed by the amplitude of the power spectrum  in Eq.~\eqref{aoergniujnqweed}. For $N=68$, using Eq.~\eqref{aogbiwaer}, we obtain 
\begin{eqnarray}
    \frac{m^2}{\M^2}\in [2.518,2.773]\times10^{-11},
    \label{aoeabnwef}
\end{eqnarray}
so that \mbox{$m\sim 10^{-11/2}m_{\rm P}\sim 10^{13}\,$GeV}.
This range of values is in agreement with what was obtained in Fig. \ref{apihdhbfhuabsdfasdf}.

Lastly, we have already obtained (see Eq.~\eqref{aojdjgbiawer}) that as long as
\begin{eqnarray}
    \alpha>8.7\times10^7
\end{eqnarray}
the observational bound $r<0.056$ is satisfied. Indeed, using the obtained values for $N$ and $m^2$ and the lower bound for $\alpha$ in Eq.~\eqref{aoijgbwe} gives 
\begin{eqnarray}
    r\in[0.050,0.053],
\end{eqnarray}
which is within  observational bounds, as expected.

The results obtained in this subsection are summarized in the $r-n_s$ graph in Fig.~\ref{qripgbuqerf}.

We can also obtain the displacement of the canonical field in the Jordan frame $\varphi$, as was discussed at the end of Sec. \ref{aosdbfaisbdfiasdf}. Using Eq. \eqref{adofbabsihdf} with the obtained value for $m^2/\M^2$, the displacement of the inflaton field $\Delta \phi\sim 0.7\M$, in the limit when $\alpha\sim10^{10}$ (represented by the yellow color in Fig. \ref{qripgbuqerf}) corresponds to
\begin{equation}
    \Delta \varphi \sim 0.7\M.
\end{equation}
In this limit $\Delta \varphi$ behaves as $\Delta \phi$, in the sense that for arbitrarily large $\alpha$, $\Delta \varphi$ becomes arbitrarily small.
We conclude that in this regime, the potential $V(\varphi)=m^2\varphi^2/2$ belongs to the small-field class of inflationary models.

In the opposite regime, when $\alpha$ takes a value around its lower bound $\alpha\sim 10^8$, the displacement of the inflaton $\Delta \phi\sim6\M$ (cf. Eq. \eqref{adsofiabnsdfasdf}) corresponds to
\begin{equation}
    \Delta \varphi \sim 6\M.
\end{equation}

To end this subsection, we can verify that the approximations made above are valid. With the obtained values for $m^2$ and $\alpha$, the value $x_{\text{end}}$ at the end of inflation is
\begin{eqnarray}
  x_{\text{end}}^2=4\frac{m^2}{\M^2}\alpha=0.0091 \Rightarrow x_{\text{end}}=0.095,
\end{eqnarray}
and the approximation made in Eq. \eqref{aogiubwefd} is valid.

Finally, the potential \eqref{aifugfbaniwef} with the obtained values of $m^2$ and $\alpha$ is
\begin{equation}
    \bar{V}(x_{\text{end}})=\frac{m^2\M^2}{1+4\alpha\frac{m^2}{\M^2}}\simeq m^2\M^2\sim 2.5\times10^{-11}\M^4,
\end{equation}
which is similar to the typical inflationary energy scale $V\sim 10^{-13}\M^4$ and in the last step we used Eq. \eqref{adofasiudfhasdf}.

\subsection{$n=4$}

In this section we focus on the $n=4$ case, following the same steps as in the last subsection. The potential in the Jordan frame, remembering $\varphi\gg M$ during inflation, reads
\begin{eqnarray}
    V(\varphi)=\lambda^4 \varphi^4.
\end{eqnarray}

Choosing $n=4$ in Eqs. \eqref{aibgiWE}-\eqref{aigbaerg}, the inflationary observables now read
\begin{eqnarray}
    A_s=\frac{8}{3\pi^2}\lambda^4(N+1)^3,
    \label{adofubawef}
\end{eqnarray}
\begin{eqnarray}
    n_s=1-\frac{3}{N+1},
    \label{aouabiuwdf}
\end{eqnarray}
and
\begin{eqnarray}
    r=\frac{16}{(N+1)\left[1+256\alpha\lambda^4(N+1)^2\right]},
    \label{aodfabdf}
\end{eqnarray}
where $N=60+\Delta N$ is the total number of inflationary e-folds. Furthermore, Eq. \eqref{aogiubwefd} now reads
\begin{eqnarray}
    x_{\text{end}}^2=16\lambda^2\sqrt{\alpha},
\end{eqnarray}
while the increase in the number of e-folds is 
\begin{eqnarray}
    \Delta N=2+\frac{1}{4}\ln{\left(\frac{1+256\alpha \lambda^4}{64 \lambda^4}\right)}.
    \label{aeoiringoaerff}
\end{eqnarray}
The above is reduced to $\Delta N=2-\ln(2\sqrt 2\lambda)$ when \mbox{$|x_{\rm end}|\ll 1$}.

In order to obtain the most accurate value for $\Delta N$, one can solve for $\lambda ^4$ in Eq.~\eqref{adofubawef} and use it in Eq.~\eqref{aeoiringoaerff} to obtain the equation
\begin{eqnarray}
    \Delta N=2+\frac{1}{4}\ln{\left[\frac{(61+\Delta N)^3+96\alpha\pi^2 A_s}{24\pi^2A_s}\right]}.
\end{eqnarray}

Using the lower bound for alpha $\alpha \sim 1.18\times 10^8$, given by Eq.~\eqref{aojdjgbiawer}, and the observational value for the amplitude of the scalar power spectrum, given by Eq.~\eqref{aoergniujnqweed}, this equation can be numerically solved to obtain 
\begin{eqnarray}
    \Delta N=8.825\simeq 9,
\end{eqnarray}
which means that the total number of inflationary e-folds is 
\begin{eqnarray}
    N\simeq 69.
    \label{adououbabdf}
\end{eqnarray}

Using this result in Eq.~\eqref{aouabiuwdf} immediately gives 
\begin{eqnarray}
    n_s=0.9571,
\end{eqnarray}
which is outside the $1\sigma$ bounds in Eq.~\eqref{aoidfbiadf} but could be accommodated by the $2\sigma$ bounds in Eq.~\eqref{aodgbuawefde}.

Using the number of e-folds in Eq.~\eqref{adououbabdf} and the observational value for the amplitude of the scalar power spectrum in Eq.~\eqref{aoergniujnqweed}, it follows from Eq.~\eqref{adofubawef} that the value the coupling constant takes is
\begin{eqnarray}
    \lambda^4\in [2.153,2.371]\times 10^{-14}.
\end{eqnarray}

This range of values is in agreement with what was obtained in Fig. \ref{aajewfaibwdfhauwefb}.

As for the parameter $\alpha$, we have already obtained (see Eq. \eqref{aojdjgbiawer}) that as long as 
\begin{eqnarray}
    \alpha>1.18\times10^8,
    \label{aeorjgnwed}
\end{eqnarray}
the bound $r<0.056$ is satisfied. Indeed, using the obtained values for $N$, $\lambda^4$ and the lower bound for $\alpha$ in Eq.~\eqref{aodfabdf} gives 
\begin{eqnarray}
    r\in [0.0507,0.0546],
\end{eqnarray}
which is within observational bounds, as expected. 

The results obtained in this subsection are summarized in the $r-n_s$ graph in Fig. \ref{qpiuerbgbhubqaedf}.

With these values for $\lambda^4$ and $\alpha$, the value $x_{\text{end}}$ at the end of inflation is
\begin{eqnarray}
    x_{\text{end}}^2=16\lambda^2\sqrt{\alpha}=0.026 \Rightarrow x_{\text{end}}=0.16,
\end{eqnarray}
and the approximation made in Eq. \eqref{aogiubwefd} is valid.

Finally, the potential \eqref{aifugfbaniwef} with the obtained values of $\lambda^4$ and $\alpha$ is
\begin{equation}
    \bar{V}(x_{\text{end}})=\frac{64\lambda^4\M^4}{1+256\alpha\lambda^4}\simeq 64\lambda^4\M^4\sim 10^{-12}\M^4,
\end{equation}
which is similar to the typical value of the inflationary energy scale $V\sim 10^{-13}\M^4$ and in the last step we used Eq. \eqref{adofasiudfhasdf}.

It is important to emphasize that the results obtained above are indicative only. The parameter $n$ can assume other order unity values, for example \mbox{$n=1$} and \mbox{$n=3$}, or even non-integer values inbetween. In Figs.~\ref{oareygboaebrf} and \ref{airbahuebrfwwww} the cases \mbox{$n=1$} and \mbox{$n=3$} are also considered. We find that the best results are obtained for \mbox{$n\simeq 2-3$}, which suggests that modelling the inflationary plateau as a power-law is a successful choice.

\section{Quintessential Sector}

We have already analysed inflation and kination in this model. In this section we focus on the positive branch of the modified Peebles-Vilenkin potential in Eq.~\eqref{aouguan} to study quintessence.

The kinetic term in the action \eqref{aoiodfgjbairg} for the field $\varphi$ in the Einstein frame, at large field values $\varphi\gg M$, reads
\begin{eqnarray}
    \frac{\frac{1}{2}(\bar{\nabla}\varphi)^2}{1+\frac{4\alpha}{\M^4}V(\varphi)}\simeq\frac{\frac{1}{2}(\bar{\nabla}\varphi)^2}{1+\frac{4\alpha\lambda^n}{\M^n}\frac{M^{n+q}}{\varphi^q}}.
\end{eqnarray}

It can be made canonical by means of the transformation
\begin{equation}
    \text{d}\phi=\frac{\text{d}\varphi}{\sqrt{1+\frac{4\alpha\lambda^n}{\M^n}\frac{M^{n+q}}{\varphi^q}}}=\left(\frac{4\alpha\lambda^n M^{n+q}}{\M^n}\right)^{1/q}\frac{\text{d}y}{\sqrt{1+y^{-q}}},
    \label{arugaiuweqwe}
\end{equation}
where we have defined 
\begin{eqnarray}
    y\equiv \left(\frac{\M^n}{4\alpha\lambda^n M^{n+q}}\right)^{1/q}\varphi,
    \label{aojbaiwef}
\end{eqnarray}
and $\phi$ can be identified as the quintessence field, or, in other words, as the inflaton field at large positive values in field space. 

The potential in the Einstein frame reads
\begin{eqnarray}
    \bar{V}&&=\frac{V(\varphi)}{1+\frac{4\alpha}{\M^4}V(\varphi)}=\frac{\lambda^nM^{n+q}/\M^{n-4}\varphi^q}{1+\frac{4\alpha\lambda^n}{\M^n}\frac{M^{n+q}}{\varphi^q}}\nonumber\\
    &&=\frac{\M^4}{4\alpha}\frac{y^{-q}(\phi)}{1+y^{-q}(\phi)}=\frac{\M^4}{4\alpha}\frac{1}{y^q(\phi)+1}.
    \label{arogbwier}
\end{eqnarray}

Note that in order to obtain an expression of the potential in terms of the inflaton $\bar{V}(\phi)$ we need to solve Eq. \eqref{arugaiuweqwe} to obtain $y=y(\phi)$ and then plug this result in Eq. \eqref{arogbwier}.

\subsection{Corrections Coming From the Matter Action}

In this section we study the influence of the coupling between the inflaton and the matter action in the Einstein frame (\textit{cf.} Eq. \eqref{aoiodfgjbairg}), following the results obtained in Sec. \ref{aidbfbiahsdf}. After making the field redefinition given by Eq. \eqref{arugaiuweqwe}, the equation of motion for the inflaton reads, using Eqs. \eqref{aosdbfbiuasdf}, \eqref{adofbasdfasdf} and \eqref{adofbbasdf},
\begin{equation}
    \ddot{\phi}+3\bar{H}\dot{\phi}+\bar{V}'(\phi)+\frac{\text{d}\varphi}{\text{d} \phi}\frac{2\alpha}{\M^4}\frac{\partial V(\varphi)}{\partial\varphi}\frac{\bar{\rho}_{\text{m}}}{1+\frac{4\alpha}{\M^4}V(\varphi)}=0,
\end{equation}
where we have taken into account that during this era $\bar{w}=0$. Using Eq. \eqref{arugaiuweqwe}, this equation can be recast as
\begin{equation}
  \ddot{\phi}+3\bar{H}\dot{\phi}+\bar{V}'(\phi)+\frac{2\alpha \bar{\rho}_{\text{m}}}{\M^4}\frac{1}{\sqrt{1+\frac{4\alpha}{\M^4}V(\varphi)}}
  \frac{\partial V(\varphi)}{\partial\varphi}=0.
    \label{aodfbbaisdf}
\end{equation}

Furthermore, the third term on the left-hand-side can be written as, using again Eqs. \eqref{arugaiuweqwe} and \eqref{arogbwier},
\begin{eqnarray}
  &&\bar{V}'(\phi(\varphi))=\frac{\text{d}\varphi}{\text{d}\phi}
  \frac{\partial\bar V(\varphi)}{\partial\varphi}\nonumber
  =\sqrt{1+\frac{4\alpha}{\M^4}V(\varphi)}\,
  \frac{\partial V(\varphi)}{\partial\varphi}\nonumber\\
    &&\times\left(\frac{1}{1+\frac{4\alpha}{\M^4}V(\varphi)}-\frac{4\alpha}{\M^4}\frac{V(\varphi)}{\left[1+\frac{4\alpha}{\M^4}V(\varphi)\right]^2}\right).
    \label{asodfasdf}
\end{eqnarray}

Putting everything together, Eq. \eqref{aodfbbaisdf} now reads
\begin{eqnarray}
  &&\ddot{\phi}+3\bar{H}\dot{\phi}+\left(1+\frac{2\alpha\bar{\rho}_{\text{m}}}{\M^4}\right)\frac{1}{\sqrt{1+\frac{4\alpha}{\M^4}V(\varphi)}}
  \frac{\partial V(\varphi)}{\partial\varphi}\nonumber\\
        &&-\frac{4\alpha}{\M^4}\frac{V(\varphi)}{\left[1+\frac{4\alpha}{\M^4}V(\varphi)\right]^{3/2}}\frac{\partial V(\varphi)}{\partial\varphi}=0
        \label{aosdifnasdf}
\end{eqnarray}

The second term inside the parenthesis, coming from the coupling of the inflaton in the matter action in the Einstein frame is Planck suppressed and, unless $\alpha$ is unrealistically large\footnote{$\alpha>\frac{m_{\rm P}^4}{\bar\rho_{\rm m}}\gtrsim\left(\frac{m_{\rm P}}{1\,{\rm eV}}\right)^4\sim 10^{108}$.}, is many orders of magnitude smaller than unity (see below the discussion concerning Eq. \eqref{aogbabiwerq} in relation to experimental constraints). Thus, the equation of motion for the inflaton during the quitessence era reads
\begin{eqnarray}
    \ddot{\phi}+3\bar{H}\dot{\phi}+\bar{V}'(\phi)\simeq 0,
\end{eqnarray}
where we have used Eq. \eqref{asodfasdf} to combine back together the derivatives of $V(\varphi)$.

We conclude the coupling in the matter action is negligible during the quintessence era and is ignored in what follows. Furthermore, note that this conclusion also holds for the matter dominated era. Indeed, the difference between both eras is that during the matter dominated era the matter energy density is the dominant contribution to the total energy density of the Universe, while during the quintessence era it is a subdominant component (accounting for $\sim 30\%$ of the total energy density). However, in both cases $\bar{w}=0$, whether the energy density of the quintessence field dominates the Universe or not, and the second term in the parenthesis in Eq. \eqref{aosdifnasdf} is negligible in both cases. Furthermore, during kination and during the radiation dominated era $\bar{w}=1/3$, so that the coupling term (given by Eq. \eqref{adofbbasdf}) vanishes. Lastly, $S_m[g_{\mu\nu},\psi]=0$ during inflation. Thus, the non-minimal coupling with the inflaton in the matter action in the Einstein frame does not affect the dynamics of the inflaton throughout the whole cosmological history of the Universe.

As for the Friedmann equation in the Einstein frame, remembering $R=-T/\M^2$ from the trace equation \eqref{aworugbiqwe}, it is easy to show that Eq. \eqref{aoijdbfaiusdf} takes the form
\begin{eqnarray}
    3\bar{H}^2\M^2&&=T_{00}+\frac{\alpha T}{\M^4}\left(T_{00}+\frac{T}{4}\right)+\frac{\alpha^2T^3}{2\M^8},
\end{eqnarray}
where
\begin{eqnarray}
    T_{00}=\frac{1}{2}\dot{\phi}^2+\bar{V}(\phi)+\rho_{\text{m}}
\end{eqnarray}
and
\begin{eqnarray}
    T=\dot{\phi}^2-4\bar{V}(\phi)-\rho_{\text{m}}.
\end{eqnarray}
Remember barred quantities are calculated using the metric in the Einstein frame \eqref{aiffgjbaierg} and dots represent $\text{d}/\text{d}\bar{t}$.

Working to first order in $\mathcal{O}(1/\M^2)$, the Friedmann equation reads
\begin{equation}
    3\bar{H}^2\M^2\simeq T_{00}=\frac{1}{2}\dot{\phi}^2+\bar{V}(\phi)+\rho_{\text{m}}\simeq \bar{V}(\phi)+\rho_{\text{m}},
\end{equation}
where in the last step we have taken into account that we work with thawing quintessence and the scalar field is only starting to roll down its potential today. 

Thus, the new effective matter sources that appear due to the treatment of our $f(R)$ function in the Palatini formalism (the terms proportional to powers of $\alpha$) are negligible compared to $T_{00}$ unless $\alpha$ is unrealistically large, and the usual Friedmann equation is recovered.

\subsection{Frozen Inflaton}

In this section we calculate the value at which the canonically normalized field $\phi$ freezes after the period of kination. It is important to mention that, although there exist other reheating mechanisms, such as instant preheating \citep{Felder:1998vq,Dimopoulos:2017tud}, curvaton reheating \citep{Feng:2002nb,Lyth:2001nq,BuenoSanchez:2007jxm}, Ricci reheating \citep{Dimopoulos:2018wfg,Opferkuch:2019zbd} or considering warm quintessential inflation \citep{Dimopoulos:2017zvq,Rosa:2019jci,Gangopadhyay:2020bxn}, in the present work we consider gravitational reheating \citep{Ford:1986sy,Chun:2009yu,deHaro:2019oki}. The reason is twofold. First, it simplifies the calculations and allows for the reader to have a clearer picture of the mechanisms behind quintessential inflation in Palatini $f(R)$ gravity. Second, this reheating mechanism propels the field the furthest after kination, so that it freezes at a value such that the residual potential energy easily fits the observed vacuum energy density. Note that gravitational reheating corresponds to the lowest possible value for $T_{\text{reh}}$, so that the increment in the number of e-folds given by Eq. \eqref{oabergiaer} is maximised. In this way, other reheating mechanism would correspond to a lower value of $\Delta N$ and, specifically, the results obtained for $n=2$ would be closer to the $1\sigma$ bounds for the scalar spectral index (see Eqs. \eqref{adfibadf}-\eqref{adofaibdfawfd}).

As it was found above (see Eqs. \eqref{aodbabhidf}-\eqref{adbabbidfa}), the equations of motion during kination read
\begin{eqnarray}
    \Ddot{\phi}+3\bar{H}\Dot{\phi}= 0,
    \label{aogbbiawer}
\end{eqnarray}
where
\begin{eqnarray}
    \bar{H}^2=\frac{\rho_\phi}{3\M^2}=\frac{\frac{1}{2}\Dot{\phi}^2}{3\M^2}.
\end{eqnarray}

This can be solved, by making the reasonable assumption (remember $\rho_{\phi}\propto a^{-6}$) that $\Dot{\phi}(t)\ll\Dot{\phi}_{\text{end}}$ when $t\gg t_{\text{end}}$, to obtain 
\begin{eqnarray}
    \phi(t)=\phi_{\text{end}}+\sqrt{\frac{2}{3}}\M \ln{\left(\frac{t}{t_{\text{end}}}\right)},
    \label{aogbiawe}
\end{eqnarray}
where $\phi_{\text{end}}$ is the value the inflaton takes at the end of inflation. At some point (at reheating) radiation takes over. Then, even though Eq. \eqref{aogbbiawer} continues to hold, the Hubble parameter becomes $\bar{H}=1/(2t)$. Solving its equation of motion during this epoch, the evolution of the inflaton reads
\begin{eqnarray}
    \phi(t)=\phi_{\text{reh}}+2\sqrt{\frac{2}{3}}\M\left(1-\sqrt{\frac{t_{\text{reh}}}{t}}\right).
\end{eqnarray}

It follows that for late times $t\gg t_{\text{end}}$ the inflaton is frozen at 
\begin{eqnarray}
    \phi_F=\phi_{\text{reh}}+2\sqrt{\frac{2}{3}}\M.
\end{eqnarray}

We can obtain $\phi_{\text{reh}}$ by evaluating Eq. \eqref{aogbiawe} at reheating and at the moment at which radiation is created, which, in the case of gravitational reheating, is at the end of inflation. Thus,
\begin{eqnarray}
    \phi_{\text{reh}}=\phi_{\text{end}}+\sqrt{\frac{2}{3}}\M\ln{\left(\frac{t_{\text{reh}}}{t_{\text{end}}}\right)},
\end{eqnarray}
so that 
\begin{eqnarray}
    \phi_F=\phi_{\text{end}}+\sqrt{\frac{2}{3}}\M\left(2+\ln{\left(\frac{t_{\text{reh}}}{t_{\text{end}}}\right)}\right).
    \label{aofojgibiabweff}
\end{eqnarray}

The ratio $t_{\text{reh}}/t_{\text{end}}$ can be estimated as follows. First, note that radiation scales as $\rho_r\propto a^{-4}$ while the background density during kination scales as $\rho\propto a^{-6}$ so that
\begin{eqnarray}
    \Omega_r=\frac{\rho_r}{\rho}=a^2. 
\end{eqnarray}

Furthermore, during kination (see Eq. \eqref{aoubbiawedf}) $a\propto t^{1/3}$. Thus, taking into account that radiation is the dominant contribution to the energy density budget at reheating, we have
\begin{eqnarray}
    &&1=\Omega_r^{\text{reh}}=\Omega_r^{\text{end}}\left(\frac{a_{\text{reh}}}{a_{\text{end}}}\right)^2=\Omega_r^{\text{end}}\left(\frac{t_{\text{reh}}}{t_{\text{end}}}\right)^{2/3}\nonumber\\
    &&\Rightarrow \frac{t_{\text{reh}}}{t_{\text{end}}}=(\Omega_r^{\text{end}})^{-3/2}.
\end{eqnarray}

Plugging this result in Eq. \eqref{aofojgibiabweff} gives
\begin{eqnarray}
    \phi_F=\phi_{\text{end}}+\sqrt{\frac{2}{3}}\M\left(2-\frac{3}{2}\ln{\Omega_r^{\text{end}}}\right).
    \label{a9urbguWAE}
\end{eqnarray}

In order to obtain an expression for the radiation density parameter at the end of inflation $\Omega_r^{\text{end}}$ we remember that the density of particles created by the event horizon in de Sitter space at the end of inflation reads
\begin{eqnarray}
    \rho_{\text{r}}^{\text{end}}=q\frac{\pi^2}{30}g_{*}^{\text{gr}}\left(\frac{H_{\text{end}}}{2\pi}\right)^4\sim 10^{-2}H_{\text{end}}^4,
\end{eqnarray}
where $q\sim1$ and $g_{*}^{\text{gr}}=\mathcal{O}(100)$ is the effective relativistic degrees of freedom. Dividing this expression by the Friedmann equation $\rho^{\text{end}}=3H_{\text{end}}^2\M^2$ gives
\begin{equation}
    \Omega_r^{\text{end}}=\frac{\rho_r^{\text{end}}}{\rho^{\text{end}}}\sim 10^{-2}\left(\frac{H_{\text{end}}}{\M}\right)^2\sim 10^{-2}\frac{V(\phi_{\text{end}})}{\M^4},
    \label{agubiauwe}
\end{equation}
where in the last step we assumed that the slow-roll approximation is valid at the end of inflation. Plugging Eq. \eqref{agubiauwe} in Eq. \eqref{a9urbguWAE} finally gives
\begin{equation}
    \phi_F=\phi_{\text{end}}+\sqrt{\frac{2}{3}}\M\left[2+3\ln{10}-\frac{3}{2}\ln{\left(\frac{V(\phi_{\text{end}})}{\M^4}\right)}\right].
\end{equation}

Using the obtained an expression for $V(\phi_{\text{end}})$ given by Eq. \eqref{aifugfbaniwef} we have
\begin{eqnarray}
    \phi_F&&=\phi_{\text{end}}+\sqrt{\frac{2}{3}}\M\bigg[2+3\ln{10}\nonumber\\
    &&-\frac{3}{2}\ln{\bigg(\frac{ n^n \lambda^n}{2^{n/2}+4\alpha n^n\lambda^n}\bigg)}\bigg].
    \label{aoregbiawef}
\end{eqnarray}

When $\alpha$ takes a value close to its lower bound, using Eqs. \eqref{adofasiudfhasdf} and \eqref{a9udsbfasdf}, this equation is simplified as
\begin{equation}
  \phi_F=\phi_{\text{end}}+\sqrt{\frac{2}{3}}\M\left[2+3\ln{10}+
    \frac{3n}{2}\ln\left(\frac{\sqrt 2}{n\lambda}\right)\right].
\end{equation}

Note that in order to obtain $\phi_{\text{end}}$ we need to solve the (generally complicated) integral \eqref{aourubgiuawef} and plug the resulting $x=x(\phi)$ in the equation for $x_{\text{end}}$ given by \eqref{aogiubwefd}. However, in most cases $\phi_{\text{end}}$ is negligible compared to the second term in the right-hand-side of Eq. \eqref{aoregbiawef}. To illustrate this we can choose the simplest case for which Eq. \eqref{aourubgiuawef} can be solved, \textit{i.e.}, for $n=2$. Indeed,
\begin{equation}
    \text{d}\phi=\frac{\M}{\lambda(4\alpha)^{1/2}}\frac{\text{d}x}{\sqrt{1+x^2}}\Rightarrow \phi=\frac{\M}{\lambda(4\alpha)^{1/2}} \sinh^{-1}{x}.
\end{equation}

Thus, 
\begin{eqnarray}
    \phi_{\text{end}}&&=\frac{\M}{\lambda(4\alpha)^{1/2}}\sinh^{-1}{x_{\text{end}}}\nonumber\\
    &&\simeq \frac{\M}{\lambda(4\alpha)^{1/2}} x_{\text{end}}
    =\sqrt{2}\M,
\end{eqnarray}
where we have used Eq. \eqref{aogiubwefd} and taken into account that unless $\alpha \gtrsim 10^{11}$, $|x_{\text{end}}|\ll 1$. Then, remembering (see Eq. \eqref{aoeabnwef}) that inflation fixes $2\lambda^2=m^2/\M^2\sim 10^{-11}$ and taking $\alpha\sim 10^8$, the inflaton freezes at 
\begin{eqnarray}
    \phi_F&&=-\sqrt{2}\M+\sqrt{\frac{2}{3}}\M\left(2+3\ln{10}+15\ln{10}\right)\nonumber\\
    &&=-\sqrt{2}\M+36\M\simeq 35 \M\gg\phi_{\rm end}. 
    \label{aorgbaiwef}
\end{eqnarray}
Notice that the above is a super-Planckian displacement of the canonical inflaton $\phi$ and not of $\varphi$, which appears in the scalar potential of this model, in Eq.~\eqref{aouguan}.

\subsection{Residual Potential Energy}

If we were to obtain the residual potential energy for a general $q$ we would need to solve Eq. \eqref{arugaiuweqwe} in order to obtain $y=y(\phi)$ and substitute it in the potential \eqref{arogbwier} to finally use the value at which the inflaton is frozen after kination, given by Eq. \eqref{aoregbiawef}. Although Eq. \eqref{arugaiuweqwe} is in general difficult to solve, we can take into account that when the inflaton stops being kinetically dominated, \textit{i.e.}, when it freezes, the potential energy has become many orders of magnitude smaller than the Plank scale (we are on the quintessential tail). In this way, we are in the regime where
\begin{equation}
    4\alpha V(\varphi)\ll \M^4 \Leftrightarrow 4\alpha \lambda^n M^{n+q}\ll \M^n\varphi^q\Leftrightarrow y^{-q}\ll 1,
    \label{aorgauiewrf}
\end{equation}
where we have used Eq. \eqref{aojbaiwef}. Thus, Eq. \eqref{arugaiuweqwe} can be approximated by 
\begin{eqnarray}
    \text{d}\phi=\left(\frac{4\alpha\lambda^n M^{n+q}}{\M^n}\right)^{1/q}\left(1-\frac{1}{2}y^{-q}\right)\text{d}y.
    \label{adiubadf}
\end{eqnarray}

This equation can be immediately integrated to obtain, for $q\neq 1$,
\begin{eqnarray}
    \phi(y)=\left(\frac{4\alpha\lambda^n M^{n+q}}{\M^n}\right)^{1/q}y\left(1+\frac{1}{2(q-1)y^q}\right).
\end{eqnarray}

Raising the above to the power of $q$ and using the approximation \eqref{aorgauiewrf} again we have 
\begin{eqnarray}
  \phi^q(y)&&
  =\frac{4\alpha\lambda^n M^{n+q}}{\M^n}\left(y^q+\frac{q}{2(q-1)}\right).
\end{eqnarray}

Therefore, the analytical expression for $y(\phi)$, in the regime defined by Eq. \eqref{aorgauiewrf}, is
\begin{eqnarray}
  y^q(\phi)=\frac{\M^n\phi^q}{4\alpha \lambda^n M^{n+q}}-\frac{q}{2(q-1)}.
    \label{aodfabisdf}
\end{eqnarray}

Evaluating this expression at $\phi_F$ and plugging it in Eq. \eqref{arogbwier}, after some algebra, we obtain the residual potential density 
\begin{eqnarray}
\frac{\bar{V}(\phi_F)}{\M^4}=\left(\frac{\M^n\phi_F^q}{\lambda^n M^{n+q}}+\frac{2\alpha (q-2)}{q-1}\right)^{-1},
  \label{ar9oughabwerf}
\end{eqnarray}
where $\phi_F$ is given by Eq. \eqref{aoregbiawef}. Note that for most values of $\alpha$, and for $q\neq1$, such that the limit $\M^n \phi_F^q\gg 2\alpha \lambda^n M^{n+q}$ holds, the potential can be approximated to first order as 
\begin{eqnarray}
    \bar{V}(\phi_F)=\frac{\lambda^n M^{n+q}}{\M^{n-4}\phi_F^q}\left[1-\frac{2(q-2)\alpha \lambda^n M^{n+q}}{(q-1)\M^n \phi_F^q}\right].
    \label{aiwengawef}
\end{eqnarray}

Also note that to zeroth order this is the same as the original Peebles-Vilenkin potential\citep{Peebles:1998qn} in the Jordan frame in the limit $\varphi\gg M$, only with $\varphi_F$ replaced by $\phi_F$. Of course, this was expected since we assumed the limit in Eq. \eqref{aorgauiewrf} in the first place.

\subsubsection{$q=1$}

Before calculating the residual potential energy density for specific values of $n$ and $q$ we focus on the special case $q=1$. Eq. \eqref{adiubadf} now reads
\begin{eqnarray}
    \text{d}\phi=\frac{4\alpha\lambda^n M^{n+1}}{\M^n}\left(1-\frac{1}{2y}\right)\text{d}y.
\end{eqnarray}

Integrating, we have
\begin{eqnarray}
    \phi=\frac{4\alpha\lambda^n M^{n+1}}{\M^n}\left(y-\frac{1}{2}\ln{y}\right).
\end{eqnarray}

It is not possible to obtain an analytic expression for $y=y(\phi)$. However, in the limit $y\gg 1$, to a good approximation
\begin{eqnarray}
    y(\phi)\simeq\frac{\M^n\,\phi}{4\alpha \lambda^n M^{n+1}},
    \label{aosdfabisdf}
\end{eqnarray}
so that the residual potential energy reads 
\begin{eqnarray}
    \bar{V}(\phi_F)\simeq \frac{\lambda^n M^{n+1}}{\M^{n-4}\phi_F}.
\end{eqnarray}

Note this coincides with the zeroth order approximation in Eq. \eqref{aiwengawef}. Of course, the approximation made in Eq. \eqref{aosdfabisdf} is equivalent to neglecting the second term in Eq. \eqref{aodfabisdf}. We can conclude that similar results to the ones obtained for a general $q$ are obtained for $q=1$.

\subsection{$q=2$ and $n=2$}

An exception for the treatment given above is $q=2$. Note that in this case the corrections in Eq. \eqref{aiwengawef} cancels out and the form of the potential for $\phi$ is the same as for the non-canonical field $\varphi$.  Furthermore, an analytical expression for $y(\phi)$ can be obtained. It reads, using $n=2$,
\begin{equation}
    \text{d}\phi=\frac{2\sqrt{\alpha}\lambda M^2}{\M}\frac{\text{d}y}{\sqrt{1+y^{-2}}}\Rightarrow \phi=\frac{2\sqrt{\alpha}\lambda M^2}{\M}\sqrt{1+y^2}.
\end{equation}

Solving for $y$ we have
\begin{eqnarray}
    y_F^2=\frac{\M^2\phi_F^2}{4\alpha\lambda^2 M^4}-1,
\end{eqnarray}
so that the potential at the value of the frozen inflaton reads
\begin{eqnarray}
    \bar{V}(\phi_F)=\frac{\M^4}{4\alpha}\frac{4\alpha \lambda^2 M^4}{\M^2\phi_F^2}=\frac{\lambda^2 M^4}{1225}\sim 10^{-14} M^4,
\end{eqnarray}
where we have used $\phi_F\simeq 35\M$ (see Eq.~\eqref{aorgbaiwef}) and that inflation fixes $2\lambda^2=m^2/\M^2\sim 2.6\times 10^{-11}$ (see Eq.~\eqref{aoeabnwef}). Note that the residual potential energy is independent of $\alpha$. 

The vacuum energy density today is $\rho_0 \sim 10^{-120}\M^4$, so that the mass scale $M$ is fixed to be 
\begin{eqnarray}
  M\sim 3.5\times 10^{-26}\M \sim
8.5\times 10^{-8} \text{GeV}.
\end{eqnarray}

\subsection{$q=4$ and $n=2$}

In this section we study the case where $q=4$ and $n=2$. We consider the lower bound $\alpha \sim 10^8$, the fact that inflation fixes $2\lambda^2=m^2/\M^2\sim 2.6\times 10^{-11}$ and the value at which the inflaton freezes $\phi_F\simeq 35 \M$. Thus, using the approximation obtained for the potential in Eq. \eqref{aiwengawef}, we have
\begin{eqnarray}
    \bar{V}(\phi_F)&&=\frac{\lambda^2 \M^2 M^6}{\phi_F^4}\left(1-\frac{4\alpha \lambda^2 M^6}{3\M^2 \phi_F^4}\right)\nonumber\\
    &&=8.7\times 10^{-18}\frac{M^6}{\M^2}\left(1- 10^{-9}\frac{M^6}{\M^6}\right).
\end{eqnarray}

The residual potential energy should be comparable to the vacuum energy density today $\rho_0\sim 10^{-120} \M^4$. In this way the mass scale $M$ is fixed by 
\begin{eqnarray}
    8.7\times 10^{-18}\frac{M^6}{\M^2}\left(1- 10^{-9}\frac{M^6}{\M^6}\right)=10^{-120}\M^4.
\end{eqnarray}

It is straightforward to solve this quadratic equation to obtain
\begin{eqnarray}
    M\sim 10^{-17} \M \sim  10 \text{GeV}.
\end{eqnarray}

\section{\label{asdifbasdf}Constraints Coming From Experimental Tests}

$f(R)$ theories in the Palatini formalism should be treated in the same way as general relativity, in the sense that they should agree with experiments and observations on all scales in order to be viable. In this way, $f(R)$ theories proposed to explain cosmic speedup should coincide with the dynamics of the solar system and laboratory experiments. In this section we summarize the most salient results found in the literature, mainly following Ref.~\citep{Olmo:2011uz}.

In scales comparable to that of the solar system the Universe does not behave as a perfect fluid (as opposed to cosmological scales), and it makes sense to make a distinction between the interior and exterior of matter sources. Outside of matter sources $\rho_{\text{m}}=0$ and, in the thawing quintessence scenario we consider, the inflaton freezes at $\phi_F$ so that $V(\phi_F)$ accounts for the vacuum energy density measured today\footnote{Remember that during the quintessence era $\phi\simeq \varphi$ to a very good approximation (\text{cf.} Eq.~\eqref{aodfabisdf}). Also, we are ignoring the fact that quintessence is thawing so, technically, it is unfreezing at present, which means that it has a non-zero kinetic energy density, which, however, is subdominant \mbox{$\frac12\dot\phi^2\ll V(\phi)\simeq V(\phi_F)$}.}. Thus, the Ricci scalar today outside of matter sources reads (\text{cf.} Eq. \eqref{oabrgribaweqq})
\begin{eqnarray}
  R_{\text{out}}\equiv R(0)=\frac{4V(\phi_F)}{\M^2}
  ={\rm constant}.
    \label{aodvabdf}
\end{eqnarray}

This means that the Einstein equations in the exterior of matter sources reduce
to the form
\begin{equation}
G_{\mu\nu}=\frac{1}{\M^2 f_R}T_{\mu\nu}-\Lambda_{\rm eff} g_{\mu\nu}\,,
\label{Ein1}
\end{equation}  
as suggested by Eq.~\eqref{aofjgbaiwner} with \mbox{$f_R(R)=\,$constant}, where
\mbox{$T_{\mu\nu}=-g_{\mu\nu}V(\phi_F)$} and
$\Lambda_{\rm eff}$ is given by Eq.~\eqref{aowrgiabuwef}
\begin{eqnarray}
  \Lambda_{\text{eff}}=\frac12R_{\text{out}}
-\frac{1}{2}\frac{f(R_{\text{out}})}{f_R(R_{\text{out}})}.
\end{eqnarray}
In the above, in view of
Eqs.~\eqref{aorauiefibnwaef}, \eqref{audfgbdfaa} and \eqref{aodvabdf} we have
\begin{eqnarray}
  f(R_{\text{out}})&&\equiv f(0)=\frac{4V(\phi_F)}{\M^2}+
  \frac{8\alpha V^2(\phi_F)}{\M^6}
 \nonumber\\
 &&= \frac{4V(\phi_F)}{\M^2}\left(1+\frac{2\alpha V(\phi_F)}{\M^4}\right),
\end{eqnarray}
and 
\begin{eqnarray}
 f_R(R_{\text{out}})\equiv f_R(0)=1+\frac{4\alpha V(\phi_F)}{\M^4}. 
    \label{aofbbaiwef}
\end{eqnarray}
Since $V(\phi_F)\simeq10^{-120}\M^4$ accounts for the vacuum energy density today and assuming that $\alpha$ is not unrealistically large, we have \mbox{$4\alpha V(\phi_F)\ll\M^4$}. Thus, the effective cosmological constant is simplified to 
\begin{eqnarray}
  \Lambda_{\text{eff}} & \simeq & \frac{2V(\phi_F)}{\M^2}\nonumber\\
&&-\frac{2V(\phi_F)}{\M^2}\left(1+\frac{2\alpha V(\phi_F)}{\M^4}\right)
  \left(1-\frac{4\alpha V(\phi_F)}{\M^4}\right)\nonumber\\
  & \simeq &
  \frac{4\alpha V^2(\phi_F)}{\M^6}\,.
  \label{ausdbfbasdf}
\end{eqnarray}

Considering the 00-component of the Einstein equations in Eq.~\eqref{Ein1} we
obtain the Friedman equation, which reads
\begin{eqnarray}
3H^2\M^2 &=& \frac{T_{00}}{f_R}+\M^2\Lambda_{\rm eff}\nonumber\\
&\simeq& V(\phi_F) \left(1-\frac{4\alpha V(\phi_F)}{\M^4}\right)+
\frac{4\alpha V^2(\phi_F)}{\M^4}\nonumber\\
&=& V(\phi_F).
\label{L}
\end{eqnarray}

Thus, the vacuum density is $V(\phi_F)$, which is much larger than
\mbox{$\M^2\Lambda_{\rm eff}$} since
\begin{equation}
\frac{V(\phi_F)}{\M^2\Lambda_{\rm eff}}=\frac{\M^4}{4\alpha V(\phi_F)}\gg 1\,.
\end{equation}  
This means that  $V(\phi_F)/\M^2$ is the ``true'' cosmological constant, as we
assumed in the previous section, while the contribution due to Palatini gravity
\mbox{$\M^2\Lambda_{\rm eff}$} is negligible. In the following we redefine
$\Lambda_{\rm eff}$ as \mbox{$\Lambda_{\rm eff}=V(\phi_F)/\M^2$}.

\subsection{Solar System}

In Sec. \ref{aoruogabiuwef} we found (see Eq. \eqref{argbaiwebf}) that the vacuum equations of motion in Palatini $f(R)$ theories are equivalent to those of GR with a cosmological constant, given by Eq. \eqref{aowrgiabuwef}.
Furthermore, we found that in the quintessential inflation scenario with the $f(R)$ function given by
\begin{eqnarray}
    f(R)=R+\frac{\alpha}{2\M^2}R^2,
\end{eqnarray}
the equations of motion are also equivalent to those of GR with a cosmological constant, now given by \mbox{$\Lambda_{\rm eff}=V(\phi_F)/\M^2$}.
It follows that, if one considers a spherically symmetric non-rotating mass distribution, such as the Sun, the metric outside is the Schwarzschild-de Sitter solution
\begin{eqnarray}
    \text{d}s^2=-A(r)\text{d}t^2+\frac{\text{d}r^2}{A(r)}+r^2\text{d}\Omega^2,
    \label{aoidbiqwe}
\end{eqnarray}
where $A(r)=1-2GM/r-\Lambda_{\rm eff} r^2/3$, with $M$ identified as the mass of the star and $\Lambda_{\rm eff}$ is the cosmological constant.
In the vacuum case, some authors \citep{Sotiriou:2005hu,Vollick:2003aw} conclude that Palatini $f(R)$ theories are compatible with solar system observations, based on the fact that for a suitable region in the parameter space of the theory $\Lambda_{\text{eff}}$ can be made small enough and predictions are virtually indistinguishable from those of the Schwarzschild solution in general relativity (which pass all experimental tests). In the quintessential inflation case, $\Lambda_{\text{eff}}=V(\varphi_F)/\M^2$ is obviously very small and the metric effectively takes the Schwarzschild form.

However, as it is pointed out in Ref.~\citep{Olmo:2011uz}, Eq. \eqref{aofjgbaiwner} departs from GR with an effective cosmological constant in the regions of space where $R$, and therefore $f_{R}$, is no longer constant (and the $\partial f_{R}$ in the right-hand-side of Eq. \eqref{aofjgbaiwner} are no longer zero), such as in the interior of stars. In this way, the transition from the interior to the exterior solution is, in general, not as simple as in GR, due to the modified dynamics in the interior of the sources.

We now give a brief overview of the study of the transition from the interior to the exterior solution in Palatini $f(R)$ theories. The reader is referred to Ref. \citep{Olmo:2011uz} for further details. It is convenient to perform a conformal transformation $g_{\mu\nu}\rightarrow h_{\mu\nu}= \gamma(T) g_{\mu\nu}\equiv \frac{f_{R}(T)}{f_{R}(0)} g_{\mu\nu}$ under which Eq. \eqref{aofjgbaiwner} reads\footnote{Note Eq. \eqref{aoidjgjbiabwe} is the same as Eq. \eqref{aisdubfasdf}, only with $\bar{g}_{\mu\nu}$ replaced by $h_{\mu\nu}$, $f_R(T)$ by $\gamma (T)$ and $\M$ by $\tilde{m}_{\text{P}}$. Indeed, $\bar{g}_{\mu\nu}=f_R(0)h_{\mu\nu}$, but the Einstein tensor is invariant under constant rescalings of the metric $G_{\mu\nu}(\bar{g})=G_{\mu\nu}(f_R(0)\bar{g})$.}
\begin{eqnarray}
    G_{\mu\nu}(h)=\frac{1}{\tilde{m}_{\text{P}}^2\gamma(T)}T_{\mu\nu}-\Tilde{\Lambda}(T)h_{\mu\nu},
    \label{aoidjgjbiabwe}
\end{eqnarray}
where we have relabelled $f_{R_{\text{out}}}\equiv f_{R}(0)$ (see Eq. \eqref{aofbbaiwef}), $\tilde{m}_{\text{P}}^2=\M^2f_{R}(0)$ and $\Tilde{\Lambda}(T)=(Rf_{R}-f)/(2f_{R}(0)\gamma^2)$, so that $\Tilde{\Lambda}(0)=\Lambda_{\text{eff}}$.

We now focus on spherically symmetric pressureless bodies, for which an analytical solution for an arbitrary $f(R)$ can be obtained \citep{Olmo:2006zu} by using the ansatz 
\begin{eqnarray}
    &&\text{d}s^2=g_{\mu\nu}\text{d}x^{\mu}\text{d}x^{\nu}=\frac{1}{\gamma(T)}h_{\mu\nu}\text{d}x^{\mu}\text{d}x^{\nu}\nonumber\\
    &&=\frac{1}{\gamma(T)}\left[-B(r)\text{e}^{2\Phi(r)}\text{d}t^2+\frac{1}{B(r)}\text{d}r^2+r^2\text{d}\Omega^2\right].
    \label{audbgiqwe}
\end{eqnarray}

The explicit form of $B(r)$ and $\Phi(r)$, obtained from the field equations \eqref{aoidjgjbiabwe}, can be found in Ref.~\citep{Olmo:2006zu}. For our current purposes it suffices to say that both functions are well defined and provide a complete solution for a nonrotating, pressureless, spherically symmetric body. Furthermore, in the exterior of matter sources, where $\gamma(0)=1$, the line element in Eq.~\eqref{audbgiqwe} is the same as the Schwarzschild-de Sitter one given by Eq. \eqref{aoidbiqwe}, just by absorbing the $\text{e}^{2\Phi}$ factor with a time coordinate redefinition and identifying $A(r)$ with $B(r)$. As for the interior of the body, the usual GR expressions are recovered by choosing $\gamma=1$ and $\Tilde{\Lambda}=0$. In this way, the Newtonian limit of the general solution \eqref{audbgiqwe} can be studied. In particular, we focus on the time-time component of the metric\footnote{We have redefined $B(r)=1-2\tilde{G}M(r)/r.$}
\begin{eqnarray}
    g_{tt}=-\frac{1}{\gamma(T)}\left[1-\frac{2\tilde{G}M(r)}{r}\right]\text{e}^{2(\Phi(r)-\Phi_0)}.
\end{eqnarray}

The conclusions presented in Ref.~\citep{Olmo:2011uz} imply that, for a Palatini $f(R)$ theory to be viable, the function $f(R)$ has to be chosen such that $\gamma(T)$ (or $f_{R}(T)$) is not very sensitive to density variations over the range of densities accessible to the corresponding experiments. In other words, $\gamma(T)$ must be almost constant since then, with a simple constant rescaling of the metric, the constant $\gamma(T)\simeq \gamma_0+\text{corrections}$ can be brought to the form $\tilde{\gamma}(T)=1+\text{corrections}$. This, in turn, implies that the metric has the standard form $g_{\mu\nu}=\eta_{\mu\nu}+\text{corrections}$. 

From a more analytical perspective, we require that a change $\Delta \gamma$ relative to $\gamma$ induced by a change $\Delta \rho$ relative to $\rho$ must be small
\begin{equation}
    \Bigr|\frac{\rho}{\gamma}\frac{\partial \gamma}{\partial \rho}\Bigr|=\Bigr|\frac{\rho}{f_{R}}\frac{\partial f_{R}}{\partial \rho}\Bigr|\ll 1.
    \label{aiajrbghbqwwwwq}
\end{equation}

This condition is equivalent to \citep{Olmo:2005zr}
\begin{equation}
    \Bigr|\frac{\rho}{\M^2R f_{R}}\Bigr|\Bigr|\frac{1}{1-f_{R}/(Rf_{RR})}\Bigr|\ll1.
\end{equation}

We now have the tools to determine whether our $f(R)$ function, given by
\begin{eqnarray}
    f(R)=R+\frac{\alpha}{2\M^2}R^2,
\end{eqnarray}
satisfies Solar System bounds or not. Using Eqs. \eqref{audfgbdfaa},
\eqref{aourugbiuwbef} and \eqref{aodbfibdsf} we have the following expressions inside the matter sources
\begin{eqnarray}
    f(T)&&=\frac{1}{\M^2}[\rho_{\text{m}}+4V(\varphi_F)]+\frac{\alpha}{2\M^6}[\rho_{\text{m}}+4V(\varphi_F)]^2\nonumber\\
    &&\simeq \frac{\rho_{\text{m}}}{\M^2}\left(1+\frac{\alpha\rho_{\text{m}}}{2\M^4}\right),
\end{eqnarray}
\begin{equation}
    f_{R}(T)=1+\frac{\alpha}{\M^4}[\rho_{\text{m}}+4V(\varphi_F)]\simeq 1+\frac{\alpha}{\M^4}\rho_{\text{m}},
\end{equation}
and
\begin{eqnarray}
    f_{RR}(T)=\frac{\alpha}{\M^2},
\end{eqnarray}
where we have taken into account that $V(\varphi_F)\ll \rho_{\text{m}}$ inside matter sources. Plugging these results in Eq. \eqref{aiajrbghbqwwwwq} gives
\begin{eqnarray}
    \frac{\alpha \rho_{\text{m}}}{\M^4}\ll 1.
    \label{aogbabiwerq}
\end{eqnarray}

It is obvious that this bound is satisfied for most values of the coupling constant $\alpha$, and in particular for the lower bounds given by Eq. \eqref{aojdjgbiawer}. For example, the density of the Sun is $\rho=1.41 \text{g}/\text{cm}^3=1.3\times 10^{-91}\M^4$, so that 
\begin{eqnarray}
  10^{-91}\alpha  \ll 1.
\end{eqnarray}

We conclude that our model passes the Solar System constraints. However, this is not the end of the story: We have overlooked one important subtlety by taking the approximation that the considered matter distributions are perfectly homogeneous. Indeed, the real structure of matter is discrete and our results could be modified. Specifically, the condition that $\gamma(T)$ has to be almost constant does not necessarily hold when one considers microscopic experiments, since it would be always possible to find regions of space where $\gamma(T)$ could take any possible value. 

\subsection{Microscopic Experiments}

In this section we make use of the results found in Refs.~\citep{Olmo:2011uz,Olmo:2008ye,Li:2008fa}. The first experimental constraint is obtained by considering the non-relativistic Schr\"{o}dinger equation for an electron in an external electromagnetic field, derived from the equation for a Dirac field in curved space-time. It is found that the $\gamma(T)$ term in the metric in Eq.~\eqref{audbgiqwe} induces a miss-match in $\tilde{m}\equiv m\gamma^{-1/2}$, where $m$ is the mass of the electron, calculated in vaccuum and in the interior of sources. This miss-match in turn corresponds to a change in the potential in the outermost part of the atom, which could induce a probability flux towards infinity reducing its half-life. In order for the miss-match to be small enough, any viable $f(R)$ theory must have a negligible \citep{Olmo:2008ye}
\begin{eqnarray}
    \Delta_m=m_0\left(\sqrt{\frac{f_{R}(\infty)}{f_{R}(0)}}-1\right),
    \label{aowebgijqwe}
\end{eqnarray}
where $m_0$ is a constant of the order of the mass of the electron $m$ and $f_{R}(\infty)$ is $f_{R}$ evaluated in the regions of space where the matter energy-density is much larger than the vacuum energy-density. From the results obtained in the previous section, we have 
\begin{eqnarray}
    \Delta_m=m_0\left(\sqrt{1+\frac{\alpha \rho_e}{\M^4}}-1\right)\simeq \frac{m_0\alpha \rho_e}{2\M^4}.
\end{eqnarray}

Since the vacuum-density scale $\M^2/\alpha$ is much larger than any matter-density scale that the wavefunction of the electron can reach, unless $\alpha$ is unrealistically large, we conclude that our choice for the $f(R)$ function is compatible with experiments related to the stability of the Hydrogen atom.

Another constraint was obtained in Ref.~\citep{Li:2008fa} from the variation in the energy levels of Hydrogen, for models in which the constraint given by Eq. \eqref{aowebgijqwe} is satisfied, \textit{i.e.}, $\Delta_m$ is negligibly small, such as ours. It reads
\begin{eqnarray}
    \Bigr|\frac{f_{RR}(0)H_0^2}{f_{R}(0)}\Bigr|\leq 4\times 10^{-40}.
\end{eqnarray}

Using the results obtained in the previous section and the first Friedmann equation we obtain
\begin{eqnarray}
 \Bigr|\frac{\alpha \rho_0}{3\M^4}\Bigr|\leq 4\times 10^{-40},
 \label{arogjbjawier}
\end{eqnarray}
where $\rho_0$ is the energy-density of the Universe today, which value is 
\begin{eqnarray}
    \rho_0\simeq 8.5 \times 10^{-30} \frac{\text{g}}{\text{cm}^3}\simeq 10^{-120}\M^4.
\end{eqnarray}

The bound in Eq.~\eqref{arogjbjawier} is obviously satisfied unless, again, $\alpha$ is unrealistically large. 

This concludes the section about constraints coming from experimental tests. We have found that our choice for $f(R)$ passes the constraints coming from both Solar System and microscopic experiments. Furthermore, they are compatible with the bound in Eq.~\eqref{aeorjgnwed} coming from inflationary dynamics. 

\section{Discussion and Conclusions}

The emphasis in this work is put on investigating quintessential inflation in
the context of an $R+R^2$ Palatini modified gravity theory. In the Palatini
formalism, $R+R^2$ gravity does not introduce an extra dynamical degree of
freedom (the scalaron) as is the case in the metric formalism. Instead,
inflation is driven by an explicitly introduced inflaton field. What the
Palatini setup does is it ``flattens'' the scalar potential leading to an
effective inflationary plateau even though the original inflaton potential might
be steep. As such, we have shown that a theory with e.g.
\mbox{$V\propto\varphi^2$} is
successful in accounting for the inflationary observables.

However, thus far this is not a new result, as inflation in the
Palatini context has been studied before. In our work we have also investigated
other implications of our Palatini modified gravity theory after inflation.
During radiation domination $R=0$, which implies that our $R+R^2$ Palatini
modified gravity does not really differ from standard Einstein gravity.
However, this is not true during kination and subsequently during the recent
history of the Universe, after the end of the radiation era. In principle,
these periods may be affected and we have studied this in detail. We have
shown that the Palatini corrections are largely subdominant to negligible
during the kination era if the the coupling $\alpha$ of the $R^2$ term
in our theory is not too large.\footnote{Recall that the Lagrangian density of
gravity is actually \mbox{${\cal L}=\frac12 m_{\rm P}^2R+\frac{1}{4}\alpha R^2$}.}
We also showed that, as far as the Universe dynamics
is concerned, the recent matter era is also unaffected.

There is an additional level on which our Palatini setup outperforms 
$R+R^2$ modified gravity theory in the metric setup, and it has to do with
constraints from experimental tests on the coupling $\alpha$ of the $R^2$ term.
The inflationary observables are satisfied when \mbox{$\alpha\gtrsim 10^8$} or
so. In the metric formalism, such values are excluded by solar system
observational constraints and other microscopic experimental tests. The tightest constraint comes from time-delay effect of the Cassini tracking for the Sun, enforcing a stringent bound on post-Newtonian parameter $|\gamma-1|<2.3\times 10^{-5}$ \citep{Hoyle:2004cw}. This implies $\alpha<5.8\times 10^{-6}$. However, this is not so
in the Palatini formalism, where experimental tests allow for large values of
$\alpha$ without problems. Thus, $R+R^2$ quintessential inflation is possible
only in the context of the Palatini and not the metric formalism.

To obtain specific predictions and demonstrate the analytic treatment of
quintessential inflation in our Palatini modified gravity theory, we have
investigated a family of models based on a generalised version of the original
Peebles-Vilenkin quintessential inflation model \citep{Peebles:1998qn},
introduced in Eq.~\eqref{aouguan}. This model is not to be taken too seriously
though. The reason is that only two small regions of the scalar potential are
really relevant. During inflation, the observable part of the scalar potential
corresponds to the region traversed in slow-roll of the canonical inflaton field
$\phi$ in no more than about 10 e-folds. For the non-canonical field $\varphi$
(cf. Eq.~\eqref{aourubgiuawef}), this region is even smaller. For thawing
quintessence, the region traversed corresponds to the field unfreezing and
starting to roll. This region is again rather small. The model approximates
the two regions as power-laws, with a positive power $n$ for inflation and a
negative power $-q$ for quintessence.

For inflation, we have shown that the correct spectral index of the primordial
curvature perturbation is obtained when \mbox{$n=2-3$}, in the case when
reheating is due to gravitational particle production. This is the
least effective mechanism for reheating, which corresponds to about
\mbox{$N\simeq 68$} e-folds of remaining inflation when the cosmological scales
exit the horizon. The problem of gravitational reheating is that the subsequent
kination period is so long that the amplification of primordial gravitational
waves challenges the process of Big Bang Nucleosynthesis. A more efficient
mechanism would reduce $N$ somewhat down to \mbox{$N\simeq 65$} or so. This
would mean that \mbox{$n\simeq 2$} or even less. The observed amplitude of the
primordial curvature perturbation determines the value of the constant
$\lambda$. When \mbox{$n=2$} we find that \mbox{$\lambda\sim 10^{-6}$}.
Finally, regarding the generated primordial tensors, we find that we 
are within the observational limits if \mbox{$\alpha\gtrsim 10^8$}.
If we are near this value, the produced primordial tensors are within reach of
observations in the near future (\textit{e.g.} by the BICEP3 or Simons observatories).

For quintessence, we have shown that coincidence can be achieved by avoiding
the extreme fine-tuning of $\Lambda$CDM. Indeed, for \mbox{$q=4$} we found
\mbox{$M\sim 10\,$GeV}, which is rather reasonable. We have shown that this
value substantially grows if $q$ becomes larger (\mbox{$M\sim 10^{-7}\,$GeV} when
\mbox{$q=2$}). However, the negative power $q$ cannot be much larger because the
barotropic parameter of thawing quintessence today would be too large
\citep{Dimopoulos:2017zvq}, the observational bound being
\mbox{$-1\leq w<-0.95$} \citep{Aghanim:2018eyx}.
Future observations (e.g. Euclid or the Nancy Grace Roman missions),
 will pinpoint $w$ further, resulting in a better estimate of $q$. It will be
 interesting if $w=-1$ was excluded and $\Lambda$CDM was in trouble.
 We should note that the power-law approximations of the scalar potential in the
 inflation and quintessence regions are only indicative. In this sense, one can
 envisage non-integer powers. 

 After inflation there is a period of kination where the inflaton field is
 oblivious of the scalar potential. Our treatment of kination within the
 Palatini setup is therefore independent of the specific model chosen for the
 scalar potential. We found that the canonical field $\phi$ is propelled over
 super-Planckian distances. However, the non-canonical field $\varphi$ for both
 inflation and quintessence (cf. Eqs.~\eqref{aourubgiuawef} and
 \eqref{arugaiuweqwe}) is expected to vary much less, as is the case of
 $\alpha$-attractors \citep{Dimopoulos:2017zvq}.
 This means that the radiative stability of the
 quintessential tail is protected and the 5th force problem of quintessence is
 overcome \citep{Dimopoulos:2017tud}.

Summing up, we have investigated quintessential inflation in the context of an
$R+R^2$ Palatini modified gravity theory. We have shown that inflation is
successful with a quadratic scalar potential for the inflaton field, while
quintessence is successful with a quartic inverse power-law potential without
the extreme fine-tuning of $\Lambda$CDM. He have found that the Palatini setup
introduces subdominant corrections to the kination and quintessence periods
and does not lead to violations on experimental tests of gravity. Our treatment
is able to provide concrete predictions for the primordial tensors and the
barotropic parameter of dark energy, which will be tested in the near future.
\begin{acknowledgments}
KD is supported (in part) by the Lancaster-Manchester-Sheffield Consortium for Fundamental Physics under STFC grant: ST/T001038/1. SSL is supported by the FST of Lancaster University. We thank Alexandros Karam for his comments.
\end{acknowledgments}

\nocite{*}

\bibliography{apssamp}

\end{document}